\documentclass[10pt,journal,compsoc]{IEEEtran}

\ifCLASSOPTIONcompsoc
  \usepackage[nocompress]{cite}
\else
  \usepackage{cite}
\fi

\ifCLASSINFOpdf
   \usepackage[pdftex]{graphicx}
\else
\fi

\usepackage{underscore}
\usepackage[document]{ragged2e}
\usepackage[ruled,vlined]{algorithm2e}
\usepackage{dblfloatfix}
\usepackage{caption}
\usepackage{hyperref}
\usepackage[all]{hypcap}
\usepackage[T1]{fontenc}
\DeclareUnicodeCharacter{0301}{\'{e}}

\begin{document}
\title{An Empirical Analysis of Implementing Enterprise Blockchain Protocols in Supply Chain Anti-Counterfeiting and Traceability}
\author{Neo~C.K.~Yiu,~\IEEEmembership{Member,~IEEE\\ Department~of~Computer~Science,~University~of~Oxford\\neo-chungkit.yiu@kellogg.ox.ac.uk}
\IEEEcompsocitemizethanks{\IEEEcompsocthanksitem Neo C.K. Yiu was with Department of Computer Science, University of Oxford, Oxford, United Kingdom}
\thanks{Manuscript first submitted for preprint on Jan 31, 2021.}}

\IEEEtitleabstractindextext{
\justify
\begin{abstract}
A variety of innovative software solutions, addressing product anti-counterfeiting and record provenance of the wider supply chain industry, have been implemented. However, these solutions have been developed with centralized system architecture which could be susceptible to malicious modifications on states of product records and various potential security attacks leading to system failure and downtime. Blockchain technology has been enabling decentralized trust with a network of distributed peer nodes to maintain consistent shared states via a decentralized consensus reached, with which an idea of developing decentralized and reliable solutions has been basing on. A Decentralized NFC-Enabled Anti-Counterfeiting System (dNAS) was therefore proposed and developed, decentralizing a legacy anti-counterfeiting system of supply chain industry utilizing enterprise blockchain protocols and enterprise consortium, to facilitate trustworthy data provenance retrieval, verification and management, as well as strengthening capability of product anti-counterfeiting and traceability in supply chain industry. The adoption of enterprise blockchain protocols and implementations has been surging in supply chain industry given its advantages in scalability, governance and compatibility with existing supply chain systems and networks, but development and adoption of decentralized solutions could also impose additional implications to supply chain integrity, in terms of security, privacy and confidentiality. In this research, an empirical analysis performed against decentralized solutions, including dNAS, summarizes the effectiveness, limitations and future opportunities of developing decentralized solutions built around existing enterprise blockchain protocols and implementations for supply chain anti-counterfeiting and traceability.
\end{abstract}

}

\maketitle
\IEEEdisplaynontitleabstractindextext
\justifying
\IEEEpeerreviewmaketitle

\ifCLASSOPTIONcompsoc
\IEEEraisesectionheading{\section{Introduction}\label{sec:introduction}}
\else
\section{Introduction}
\label{sec:introduction}
\fi

\IEEEPARstart{T}{h}e problem of counterfeit product trading has been one of the major challenges the supply chain industry has been facing, in an innovation-driven global economy. The situation has exacerbated with an exponential growth of counterfeits and pirated goods worldwide, for which it has also plagued the companies with multinational supply chain networks for decades and on, according to an analytical study – \cite{1} published back in 2016, suggested that the volume of counterfeit trading has already surpassed as much as \underline{\emph{\$509 billion}} representing \underline{\emph{3.3\%}} of world trade. Given the growing concern and worsening situation in trading activities of counterfeit products, there have been anti-counterfeiting solutions developed and implemented in supply chain systems of different industries. The motivation and research objectives, which is based on analyses on existing blockchain implementations in supply chain industry, such as the Decentralized NFC-Enabled Anti-Counterfeiting System, will be elaborated in this chapter.

\subsection{Existing Blockchain Implementations in Supply Chain Industry}
Many of the existing implementations for product anti-counterfeiting in supply chain industry including NAS as described in \cite{25}, utilizing wireless communication technologies, wireless sensor networks or geospatial technologies, are built with centralized architecture relying on trusted servers, databases and applications, which are solely controlled and managed by manufacturers of different sectors, for coordinating and managing product authentication with participation contributed from different nodes along the supply chain of different industries. As elaborated in \cite{towardblockchain}, given a variety of advantages, including prevention of single-point failure, better resilience and availability, introduced via applying blockchain technology with a concept of decentralized application, to have more secured and sophisticated supply chain systems, more and more decentralized solutions basing on a wide range of blockchain implementations and protocols have been developed and implemented in supply chain industry.

Blockchain innovations have also been implemented across supply chain industry, and some are specifically with use cases of decentralizing and improving product traceability and anti-counterfeiting aspects of the industry. \cite{55} has proposed a concept of blockchain system to enhance transparency, traceability and process integrity of manufacturing supply chains, while an Ethereum-based fully-decentralized traceability system for Agri-food supply chain management named \emph{AgriBlockIoT} was developed in \cite{56}. Furthermore, a novel blockchain-based product ownership management system, a blockchain-based anti-counterfeiting system coupled with chemical signature for additive manufacturing and ontology-driven blockchain design for supply chain provenance, are also explained in \cite{57,58,59} respectively.

Some implementations are conceptualized and developed based on computation-extensive permission-less blockchain networks and consensus algorithms, aiming at full decentralization over scalability and stability of such decentralized solutions developed. Instead of developing blockchain implementations based on conceptual design, decentralizing legacy anti-counterfeiting systems with centralized architecture already implemented in the industry, further with blockchain innovations integrated with, would be a more pragmatic way to start with. As such, the Decentralized NFC-Enabled Anti-Counterfeiting System (dNAS) is developed, as depicted in \cite{dnas}, utilizing decentralized blockchain network with enterprise blockchain protocols compatible with the concept of enterprise consortium, programmable smart contracts and a distributed file storage system to develop a secure and immutable scientific data provenance tracking and management platform, so as to provide timely support to improve the snowballing situation of product counterfeits in supply chain industry.

\subsection{Research Objectives}
Given dNAS has already been developed and implemented in supply chain industry for decentralizing anti-counterfeiting and traceability functions, with system model definitions and system architecture based on a set of fundamental system requirements defined and security analyses against its centralized counterpart, a main question of this research should therefore be:

\emph{"Are decentralized anti-counterfeiting and traceability solutions, such as dNAS, which have already been developed and implemented in supply chain industry, more effective and worth on combating the rampant counterfeiting attacks, compared with its centralized counterparts?"}

In order to progress the research with a wide range of analyses performed against dNAS to identify future opportunities of strengthening the capability of anti-counterfeiting and traceability with dNAS, the main question is then addressed through answering the following sub-questions throughout the research:
\begin{enumerate}
  \item What are the possible approaches of the integration model to such proposed decentralized solutions for which a user-friendly solution should be developed to help promote adoptions from node participants, end consumers and to supply chain industry as a whole?
  \item How is the proposed decentralized prototype tested and validated with different settings, and against its centralized counterparts qualitatively and quantitatively?
  \item What are the limitations, concerns and potential future opportunities of dNAS?
\end{enumerate}

Given the main research question and a set of the derived sub-questions identified, it is common to follow an organised way of exploring them step-wise in this research, via a background on enterprise blockchain and existing enterprise blockchain protocols applied in supply chain industry, a wide range of system analyses against the developed dNAS and a discussion of results so as to gain insights on what and how dNAS and other decentralized supply chain anti-counterfeiting and traceability solutions, basing on enterprise blockchain protocols and implementations, could further be improved.

\section{Background}
In this chapter, following blockchain fundamentals and the advantages brought by the blockchain technology and blockchain 2.0, where dNAS and many of the existing decentralized solutions in supply chain industry are currently developed on, as explained in \cite{towardblockchain}, the concept of enterprise blockchain and existing enterprise blockchain protocols currently applied in supply chain industry are also elaborated. 

\subsection{Enterprise Blockchain at the Core}
The blockchain network can generally be categorized either as permission-less (public network) or permissioned network. The former is an open distributed ledger network such as \cite{28}, where any node can join the network and where any two peers can conduct transactions without any authentication performed by any central authority, while the latter is a controlled distributed ledger like \cite{30} where the decision making and the validation process are kept to one organization or few organizations forming a consortium with or without the staking concept. In permissioned networks, the consortium administrator or certificate authority determines who can join the network as a validator node or listener node. All nodes are authenticated in advance, and their identity is known to other nodes running on the same network and in the same consortium, also known as peer-permissioning.

Permissioned blockchain implementations utilized by any organization or a group of organizations in the same subset of industry forming a consortium are commonly known and classified as "\emph{enterprise blockchain}". Different industries and service sectors have been implemented with innovative solutions and system, such as enterprise resource planning, global positioning system and wireless tag communication system, in order to enhance the effectiveness and efficiency of service processing; however these innovations are not usually integrated and seamlessly connected to form a supply chain and value chain, which should suppose to facilitate the visibility of processed data across a supply chain or a specific industry.

Enterprise blockchain implementations are \emph{not fully decentralized} compared with decentralized solutions running on public networks made available for wider audience to interact with data exchange, and it is instead more focused on building use cases to improve operational efficiency and effectiveness for a group of organizations, a subset of industry or even an entire supply chain of specific industries. Though it is not fully decentralized, it still come with the blockchain characteristics, as elaborated in \cite{towardblockchain}, and possibly a sovereign blockchain network for a group of organizations in a specific subset of industry, preventing security threats which could be identified in centralized architecture, such as single-point process, lower trust levels and sequential supply chain data exchanges amongst supply chain participants. There are more advantages and industry perceptions towards enterprise blockchain implementations which are well explained in \cite{ebc}.

\subsection{Related Work of Enterprise Blockchain Protocols}
With the advancement of blockchain development in recent years, there have been some existing blockchain innovations developed in different domains and in combination with other emerging technologies, such as IoT, NFC, artificial intelligence, machine learning and cloud computing, for different purposes like forming digital business ecosystems to propel towards Industry 4.0, as detailed in \cite{supervisedmachinelearning}. Given the advantages such as peer-permissioning, a more scalable consensus algorithms and a capability of maintaining sovereign blockchain networks for specific business use cases with specific groups of organizations forming enterprise consortia, these disruptive digital business ecosystems are in all likelihood developed basing on open-source enterprise blockchain protocols and implmentations, including Ethereum Proof-of-Authority network, Hyperledger Fabric, Consensys Quorum and Tendermint Core, as elaborated in the following.

\begin{figure*}[h]
    \centering
    \captionsetup{justification=centering}
    \includegraphics[width=0.8\textwidth]{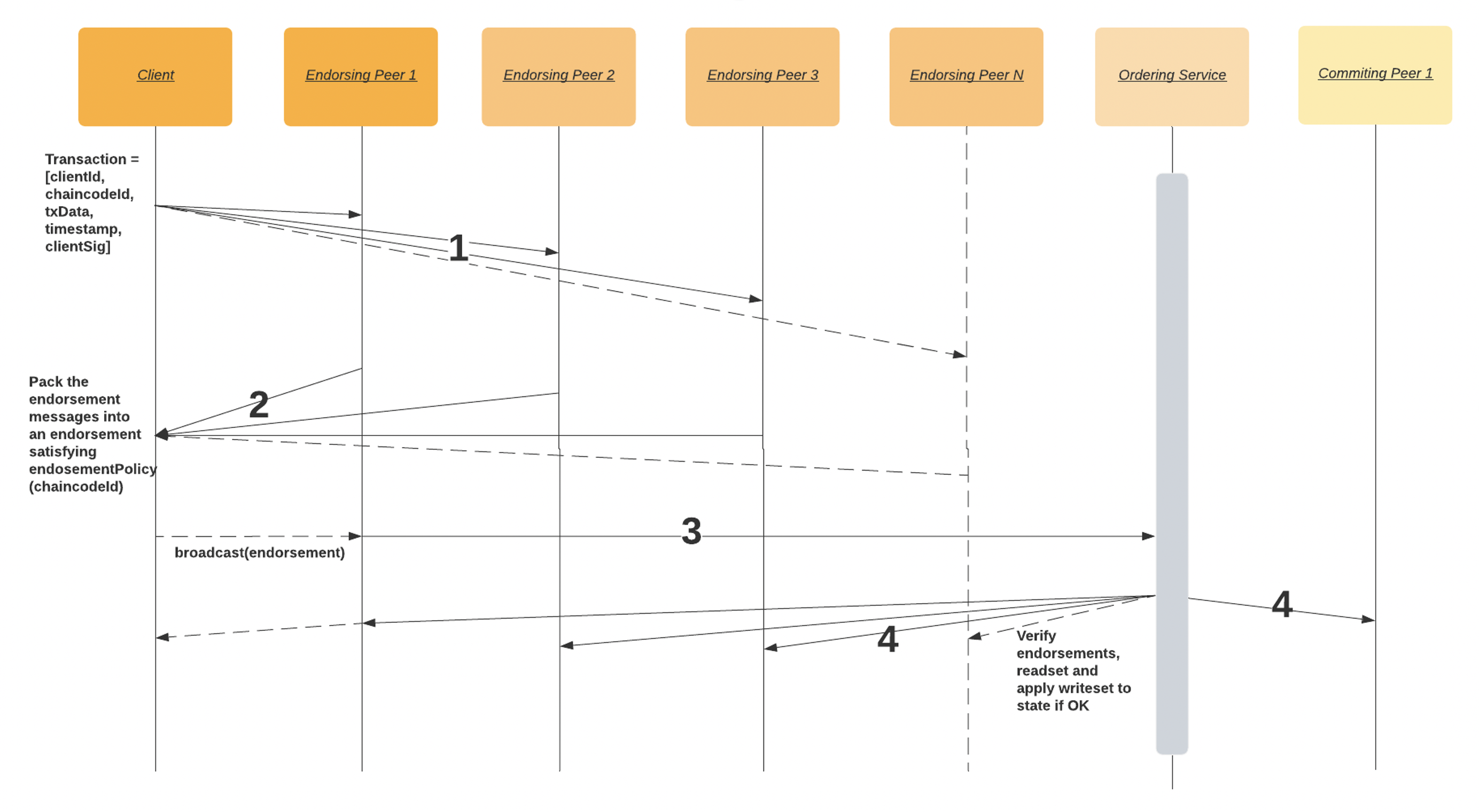}
    \caption{\textit{Hyperledger Fabric Transaction Endorsement Workflow}}
    \label{fig:hyperledgerfabric}
\end{figure*}

\subsubsection{Hyperledger Fabric}
\cite{30} is a permissioned blockchain framework implementation providing modular components designed for business use cases to build its own enterprise blockchain. It is one of the \emph{Hyperledger} projects hosted by The Linux Foundation. Hyperledger Fabric leverages container technology to host smart contracts called "\emph{chaincode}", comprised of the application logic of systems, available in a wide range of programming languages. In Hyperledger Fabric, the nodes that have access to the ledger are called "\emph{peers}", and each peer belongs to some organizations. Adding transactions to Fabric is literally a \emph{two-phase endorsement process} as shown in \textit{Fig.~\ref{fig:hyperledgerfabric}}:

\begin{enumerate}
  \item A client requests a transaction firstly approaches one or more endorsing peers with a transaction proposal and asks them to execute as well as endorsing the proposal. 
  \item The endorsing peers then execute a smart contract to determine whether or not to endorse the transaction based on an endorsement policy, and if so, endorsed transactions would then change the state on the ledger. All endorsers must see an identical transaction proposal, or else it is rejected in the \emph{ordering service}.
\end{enumerate}

The "logical validity" of transactions is determined in the endorsement phase. Once sufficiently enough endorsements are obtained, the client would then send the endorsed transaction to an ordering service, hosted by the leader node of an organization, imposing a linear order on transactions and adding them to the ledger. The number of required endorsements for a transaction is determined by an endorsement policy, which is set when the ledger is initialized. Roughly speaking, the ledger of the organization has only a single endorsement policy applied to every transaction, and there are also authentication and authorization policies applied for peer-permissioning. Sensitive data on the ledger should remain private from the chain of blocks, ordering service, and a subset of peers in a channel. Only evidence, such as hashes, need to be on the chain of blocks or sent to ordering service, later distributed to every peer of the organization.

The consensus algorithm of Hyperledger Fabric is confined in the \emph{Crash Fault Tolerant} (CFT) ordering service, aimed at fast finality, based on the \emph{RAFT} consensus protocol, which follows a "\emph{leader and follower}" model, where a leader node is elected per organization or channel, and its decisions are replicated by the followers. Such consensus algorithm is perceived to be more centralized than other consensus algorithms, amongst enterprise blockchain protocols, especially when the number of nodes on the network is low.

\subsubsection{Tendermint Core}
Tendermint Core is a blockchain application platform, developed based on \cite{42} and \cite{43}, equivalent of a web-server interaction, database, and supporting libraries for blockchain applications written in any programming language allowing creation of business task-specific blockchains fulfilling any business needs. Tendermint Core performs Byzantine Fault Tolerant (BFT) State Machine Replication (SMR) for arbitrary deterministic and finite state machines, functioning as a consensus engine for other blockchain tools, such as Hyperledger Burrow – an alternative implementation of Ethereum Virtual Machine (EVM) for the smart contracts, to be built on top of.

Given the application of \emph{leader-based BFT consensus algorithm} on Tendermint, the scalability and performance aspects of the consensus machine is greatly improved from its counterparts of the traditional Proof-of-Work. It only requires more than two-third nodes of the network to agree on a state of the ledger before a consensus is reached based on the BFT consensus.

Tendermint is a major building "\emph{block}" of the whole Cosmos Network project, also known as the Internet of Blockchains. They have implemented their decentralized Cosmos Hub on Tendermint to improve blockchain interoperability. This Hub is where all the interoperable blockchains would communicate with each other via the sidechains, also written in Tendermint, allowing transfer of digital assets from one blockchain to another. It is designed to be straightforward to integrate any ÐApp or blockchain into the Internet of Blockchains in case those are also built with Tendermint or Cosmos-SDK, which is an extension of Tendermint with added stacking.

\subsection{Contribution – System Analysis and Evaluation of dNAS}
The idea of dNAS is all about provably honest of which users of the decentralized solution should be able to automatically verify actions requested by server instances of other nodes along the supply chain, whenever they would like to and have the servers instances themselves checking with the instances of decentralized storage and decentralized network with validation results returned. dNAS aims at delivering a more secured and quality approach to verify the authenticity and provenance of luxurious products, such as bottled wine, with the use of peer-to-peer enterprise blockchain protocols basing on a concept of enterprise consortium, and IPFS which has the potential to eliminate absurd amount of costs for on-chain storage and provide a much higher level of privacy, reliability and quality of service compared with the legacy NAS of centralized architecture. 

For entities to embrace this budding technology with confidence, dNAS needs to be secured, reliable, flexible with integration models available, and its implementation details should be straightforward for supply chain nodes to adopt. Similarly, for consumers interacting with the user interface of dNAS, it needs to be user-friendly and transparent. dNAS itself does offer a foundation to explore the strengths and weaknesses of decentralised applications over their centralised counterparts, especially in supply chain industry. The evaluation of the proposed dNAS against the research objectives, and in areas, such as the decentralized wine record management, data and process integrity in dNAS, system security, availability and integration, would also uncover certain limitations and concerns of hosting and deploying ÐApps on different enterprise blockchain protocols and implementations. 

The research is aimed at investigating into the system evaluation with various system testing and performance testing activities performed against dNAS to determine if decentralized solutions applied in supply chain anti-counterfeiting and traceability are effective and useful as expected. Discussion of results is also covered to provide qualitative and quantitative system analysis, benchmarking different existing enterprise blockchain protocols to the novel dNAS developed, with risks and limitations identified for creating potential opportunities on future development and advancement of dNAS.

\section{The Decentralized NFC-Enabled Anti-Counterfeiting System (dNAS)}
dNAS is a decentralized and integrated system tailor-made for the wine industry. While capabilities such as end-to-end traceability, provenance and authenticity were enabled in the legacy NAS, as described in \cite{25}, its anti-counterfeiting model was actually based on winemaker roles with their products moving downstream to their registered supply chain participants and wine consumers until the retail points. The anti-counterfeiting model of dNAS is actually based on the decentralized enterprise consortium consisting of multiple winemakers and supply chain participants, responsible for operating the web application and NFC-enabled mobile applications of the legacy NAS on a dedicated and decentralized blockchain network, via an integrated interface, with blockchain nodes assigned to the consortium members, based on enterprise blockchain protocols.

\begin{figure*}[h]
    \centering
    \captionsetup{justification=centering}
    \includegraphics[width=1\textwidth]{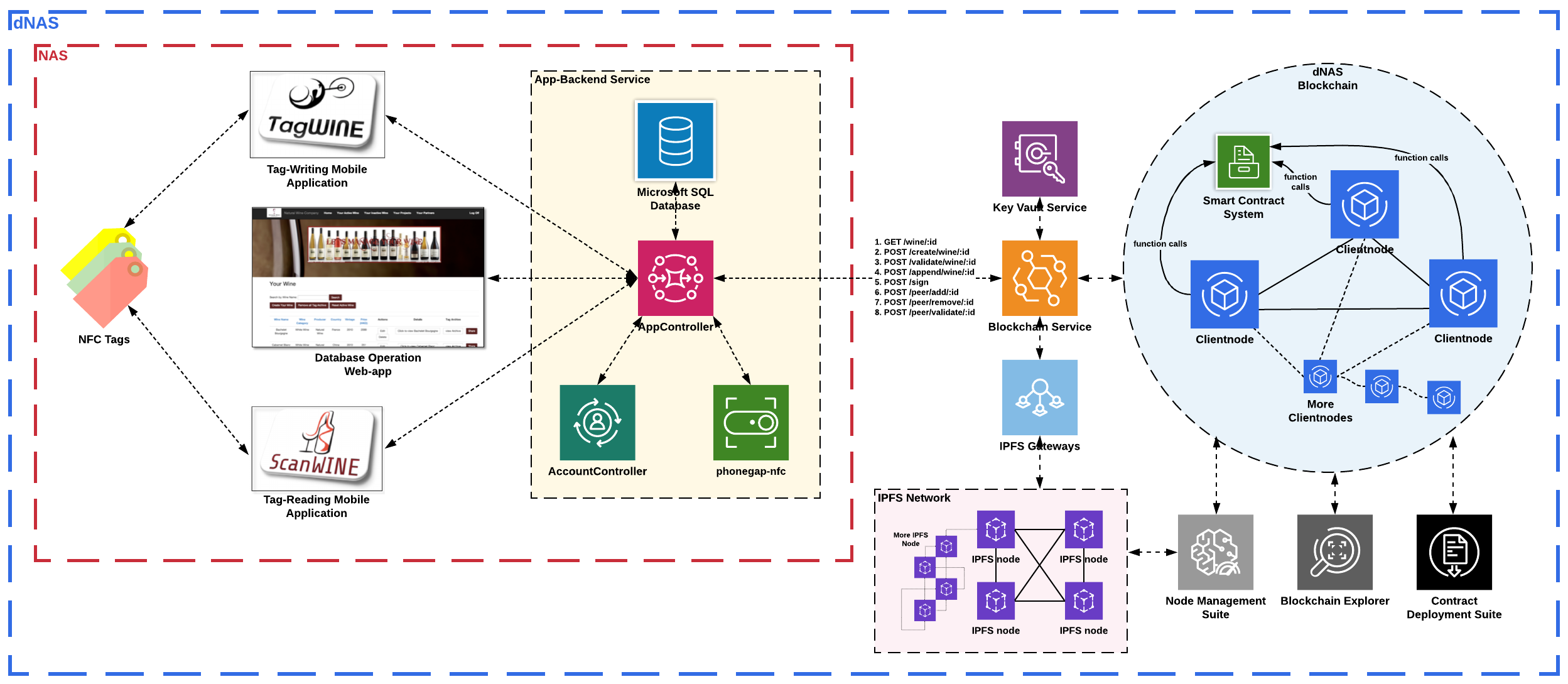}
    \caption{\textit{System Architecture Diagram of dNAS}}
    \label{fig:sysarchi}
\end{figure*}

With the proposed system architecture of dNAS as demonstrated in \textit{Fig.~\ref{fig:sysarchi}}, it explains how exactly the legacy NAS could be re-engineered and running on the chosen enterprise blockchain protocols, and what exactly are the system components to be decentralized so as to enable core functionalities of the novel dNAS. Several major design decisions were also taken into account for the system architecture implemented as expected in terms of system scalability, security and privacy. The proposed system architecture of dNAS can be explained in mainly three parts, namely (1) the decentralized blockchain network, (2) the blockchain interface, and (3) the system components built in the legacy NAS.

Given the system model design with several design decisions explained, the actual system implementation steps of dNAS were also defined, including the proposed system implementation with the system operation protocols, specified for detailing the system functionalities of the novel dNAS, on different wine record management process. The system implementation is generally categorized into \emph{two phases}, as demonstrated in \cite{dnas}, namely the initialization phase and the data processing phase, covering system implementations ranging from consortium formation, smart contract deployment to wine record management on dNAS. The initialization phase includes all the necessary system implementation steps needed to be in place before dNAS can be fully functional for wine record management when wine products moving along the supply chain. \emph{The initialization phase} is consisted of consortium formation process and smart contract deployment process. \emph{The data processing phase} includes all the necessary system operations regarding the decentralized wine record management along the supply chain, such as creation, validation and appending, involving different consortium members.

\section{System Analyses of dNAS}
System analyses are performed against system design decisions made during the development of dNAS, based on the proposed system model and system architecture of dNAS, ranging from the selection of enterprise blockchain protocols, on-chain data storage to system integration models selected for promoting adoptions of dNAS or any other existing decentralized solutions in supply chain anti-counterfeiting and traceability. 

\subsection{Selection of Blockchain Network and Consensus Protocol}
There are a variety of enterprise blockchain protocols other than the Ethereum Proof-of-Authority blockchain implementation, such as Hyperledger Fabric and Tendermint Core, for which the development of dNAS could be based on. Since trust is already achieved outside the blockchain network, the concept of enterprise consortium is therefore introduced to implement a permissioned blockchain network for every consortium member of wine industry or the wider supply chain industry, to run their blockchain nodes on a Proof-of-Authority (PoA) protocol to send and validate transactions, bypassing any intermediary acting on behalf of them. With the advent of the permissioned network based on the consortium with a concept of enterprise blockchain, non-member accounts to access methods of deployed smart contracts are then limited, based on the role restrictions applied in the smart contract design patterns. Peer-permissioning is also enabled under which every registered node must be explicitly allowed to join the network, with trusted authorities applying accountability towards the consortium formed for the wine industry. It is therefore unnecessary to commit much computational power to support the trust in an enterprise consortium entirely, allowing the consensus mechanism for enterprise solutions to be selected, instead of any computation-extensive consensus algorithm applied in public blockchain networks. It will then in turn enhance scalability and performance of the chosen blockchain network with enterprise blockchain protocols applied.

The process of achieving a non-disputable agreement, amongst the distributed nodes' states, is called consensus. The popular options of consensus algorithm for any enterprise blockchain solution are ranging from the aforementioned \emph{PoA} on Ethereum, the \emph{Crash Fault Tolerant} (CFT) algorithm such as the Reliable, Replicated, Redundant, And Fault-Tolerant (RAFT) applied in Hyperledger Fabric or Quorum, to the \emph{Practical Byzantine Fault Tolerance} (PBFT) like the \emph{Istanbul Byzantine Fault Tolerant} (IBFT) applied in Quorum Blockchain which is an enterprise blockchain platform forked from the open-source Go-Ethereum (Geth) client of Ethereum with several protocol-level enhancements to support different business needs.

\begin{figure}[h]
    \centering
    \captionsetup{justification=centering}
    \includegraphics[width=0.5\textwidth]{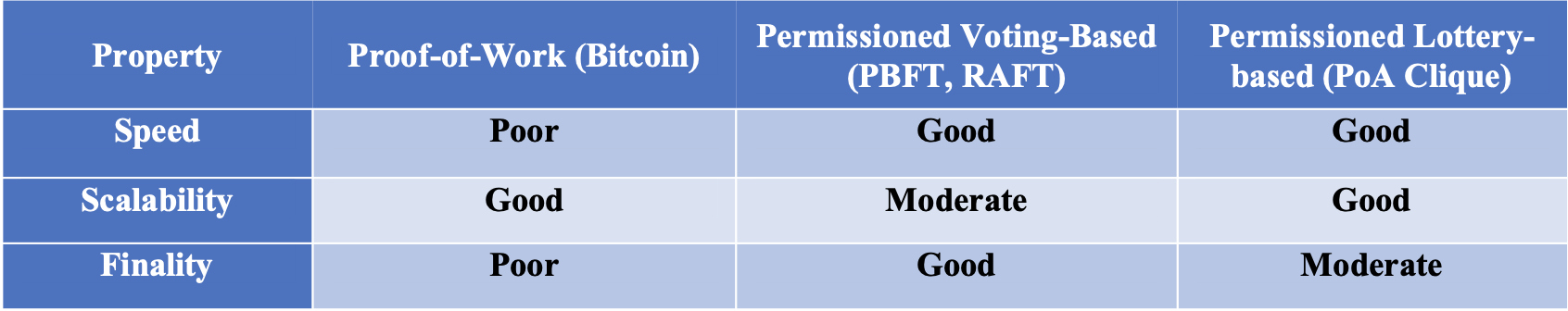}
    \caption{\textit{Comparison of Consensus Algorithms}}
    \label{fig:consensusalgocompare}
\end{figure}

Proof-of-Authority Clique, RAFT and IBFT all offer \emph{absolute finality} implying that a confirmed transaction cannot be reverted or annulled in any case. Both RAFT and IBFT, which are permissioned voting-based, offer immediate finality, where no fork is ever allowed to happen. While the permissoned lottery-based PoA Clique offers finality within the \emph{(N / 2) + 1} blocks because each of those blocks are protected by a different signer sealing blocks based on the public key cryptography, thus forming a chain of digital signatures. It is therefore achieving proof of origin and data immutability for every transaction, which cannot be broken without majority of a network colluding together. The comparison of typical consensus algorithms is listed in \textit{Fig.~\ref{fig:consensusalgocompare}}.

\emph{PoA Clique} and \emph{PoA Aura} both rely on a set of trusted authorities utilizing a simplified messaging algorithm to achieve better performance than typical PBFT algorithms. There is only one round of messages exchanged among the authorities in PoA, compared with \emph{three rounds} in PBFT, namely the \emph{pre-vote}, \emph{pre-commit} and \emph{commit} stages respectively to validate a transaction and sign a block once there are signatures amounted to \emph{two-third} of a validation group, comprised of a validation leader by vote and a group of validators. 

Therefore, better performance is one of the claims of PoA when measured against other permissioned voting-based algorithms, such as PBFT. PoA network can tolerate up to \emph{N / 2 – 1} byzantine authority nodes. It could operate correctly when \emph{N / 2 + 1} authority nodes are honest. Multiple validator nodes are allowed to propose any block. Contrary to the PBFT, the design of PoA might sacrifice consistency for better availability so faster block committal but with eventual consistency guarantee once the forks get sorted out by the \emph{GHOST} protocol. PoA produces blocks at a configurable but fixed interval, regardless of whether there is any transaction to include or not, and \emph{PoA Clique is an appropriate choice} for a network containing parties that do not trust each other. For instance, with a network of 4 blockchain nodes, 3 validator nodes and a node of consortium administrator, it would be up to \emph{N – (N / 2 + 1)} which is 1 validator node, for this case, can propose a block for any given block interval.

PoA Clique or Aura is a more appropriate choice than RAFT for the development of dNAS, given its decentralized mechanism and signature-protected data immutability of validated transaction. For instance, validator nodes on the PoA Clique network take turns to validate transactions and sign a block for which the block sealing process is lottery-based amongst the participating nodes of the network, instead of being executed by a specific leading validator node. While for RAFT, the consensus model assumed the elected leader node always acts honestly and correctly with followers blindly replicate state transitions proposed by the leader node with no further validation. In Hyperledger Fabric \cite{30}, the elected leader node controls the order of transactions endorsed in the centralized ordering service of a channel aiming at fast finality on blocks. Blocks mined in a Quorum RAFT consensus network are not protected by signatures, as a result, blockchain data can be rewritten rather easily by modifying historical data, such as transaction inputs and recalculated block hash, transactions trie root, etc. RAFT consensus does not mint new blocks unless there are pending transactions. This would result in significant storage savings, but trade-offs are difficulty and security of the blockchain which could be lowered without any signature protection. The canonical chain is far simpler than its counterparts of enterprise blockchain protocols.

\subsection{On-chain Data Storage}
Many useful applications need to persist data to be beneficial for their end users, especially if someone is considering migrating their application from a centralized approach to a decentralised alternative. Given how unscalable and how expensive it could be as it would cost absurd amount of costs in case dNAS was to build on the Ethereum main net, questions such as how, what and where the registered nodes' persistent data is stored will need to be accounted and addressed carefully.

According to a standard wine record defined dNAS, as demonstrated in \textit{Appendix~\ref{a1}} with a corresponding data model defined in \textit{Appendix~\ref{a3}}, merely one industrial operation performed with the individual transactions logged in the data schema, would already have the wine record itself sized at \emph{2,900} bytes. Given the size of the wine record calculated, should ones still look to process the whole wine record on-chain and put whole chunks under the "\emph{data}" field of every transaction, not to mention every additional supply chain process with its transaction data will at least add \emph{806} bytes to the wine record, and every additional unsuccessful validation process will at least add \emph{313} bytes to the wine record?

Looking at the gas used per transaction involving the whole \emph{2900-byte} wine record would suggest a very insightful idea on whether the whole wine record should be processed on-chain. Based on calculation steps detailed in \textit{Appendix~\ref{a4}}, total gas used per transaction and total cost for the \emph{2900-byte} wine record would be \emph{1,820,000} and \emph{0.1092 ETH} (\emph{\$133.55} with ETH priced at \emph{\$1,223}) respectively. Given the fact that dNAS is running on enterprise blockchain protocols and no gas cost is specified on any gas used in any transaction, \emph{storing the full wine record will never be a good idea} regardless of which enterprise blockchain protocol would be adopted, and regarding the performance of the blockchain network and dNAS as a whole.

It is why IPFS - a distributed file storage is adopted and deployed as a system component of dNAS connected to the blockchain service, processing a subset of a full wine record, as demonstrated in \textit{Appendix~\ref{a2}}, involved in any operation of the data processing phase, under which a standard \emph{46-byte} hash will be generated, regardless of size of the wine record for storing the data in a decentralized and distributed fashion with the original wine record to be retrieved with the generated content hash supplied to any local IPFS node. Instead of parsing a full wine record to the data field of any transaction involved with the enterprise blockchain protocol adopted in dNAS, the IPFS content hash representing a version of wine record is then stored and processed on-chain with corresponding methods of the deployed smart contracts for any blockchain transaction.

A couple of mapping data structure defined in the deployed smart contacts are based on \emph{32-byte} hash representation of the unique wine identifier as the key of mapping key-value storage so as to log the important data, such as the \emph{32-byte} hash representation of unique tag identifier, the write count (iteration) of every NFC tag, and the public address of any registered node which could be recovered from a respective signature passed in. With these types of data stored on-chain based on specific wine identifiers, on-chain states could therefore be included in any validation logic whenever a NFC tag is scanned during any accepting or buying process of specific wine products.

\subsection{Proposed Hybrid System Integration Model with Legacy NAS}
dNAS should set to provide a streamlined mechanism for registered nodes of different roles in wine industry or the wider supply chain industry, to migrate their current centralized architecture adopted in the legacy NAS to dNAS with \emph{certain degree of decentralization}, as explained in \cite{towardblockchain}. A hybrid approach should also be offered for potential nodes of winemaker role and supply chain participant role, who are less concerned about trust issues of central architecture, to join the enterprise consortium of dNAS. The hybrid approach of dNAS is a starting point for newly joined registered nodes to enjoy benefits suggested in dNAS and is more of an incremental way for any registered node in wine industry to eventually adopt the fully decentralized and distributed approach. \emph{The hybrid approach is therefore adopted} for development of dNAS depicted in \cite{dnas}.

Instead of conducting a complete architecture overhaul and having every consortium member (registered nodes of wine consumer role are not part of the consortium as mentioned in the use-case analysis performed in \cite{dnas}) hosting their own instance of dNAS, every registered node will go through use cases of registered node account management developed in \emph{AccountController} of the app-backend service, to login and perform system operations on wine records, enabled by the dNAS instances hosted by the consortium, via a trustful governing authority which is the consortium administrator. 

Having the consortium to host dNAS, it implies that every consortium member is assigned with all the distributed system components, including the blockchain service, key vault service, blockchain node, IPFS node, etc. Every registered node with access to the web application and mobile applications would require partition logic patterns applied to the database of app-backend service. In order to process and validate wine records properly, authentication logic pattern for user sessions of different applications, access control pattern to functionalities designed for different roles of registered node, and authorization logic applied to the blockchain service and app-backend service, are required to implement. The blockchain node and IPFS node could therefore be connected whenever there is a request for any interaction with the blockchain network and the IPFS network. Access to the node management suite and the blockchain explorer will need to be given to each consortium member to synchronize with the state and notify of performance on the blockchain node and IPFS node owned by them.

By having the consortium administrator to bestow inherent level of trust between consortium members, the registered nodes could access to dNAS without much hassle. With the proposed hybrid approach, despite the fact that registered nodes would get benefits of decentralisation, such as high availability and non-repudiation on the states stored on-chain, subsets of wine records stored in IPFS network and decentralized operations of the consortium, it will still be susceptible to certain vulnerabilities from the given degree of centralization attributed by the concept of enterprise consortium. The app-backend service, blockchain service and every instance of the distributed system are actually hosted by the consortium, which is managed by the consortium administrator, scenarios such as service downtime and failure on centralized system components of the app-backend service could still be applied. However, states of wine records will still be conserved within the decentralized system components of dNAS.

\section{System Testing}
A variety of software tests are in place to ensure functionalities set out in different system components are functioning as expected, the system testing process of dNAS could be divided into the functional testing and performance testing.

\subsection{Functional Testing of dNAS}
\begin{figure}[h]
    \centering
    \captionsetup{justification=centering}
    \includegraphics[width=0.45\textwidth]{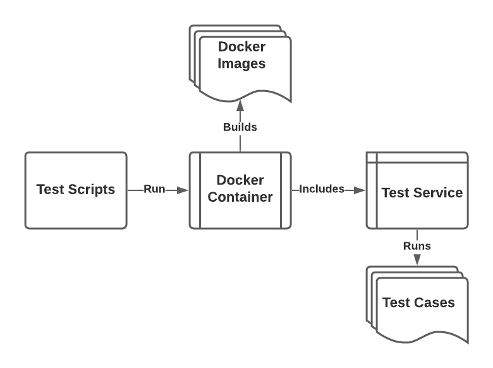}
    \caption{\textit{Test Process of dNAS}}
    \label{fig:testprocess}
\end{figure}

Concept of microservice containerization has made software testing and deployment more straightforward in dNAS. Test scripts are developed to build the corresponding containers, based on the definition of docker-compose files, which would in turn build the respective Docker images specified in its Dockerfile. The test scripts will run the specific test service of a running container, with individual test cases executed. The test process of dNAS is demonstrated in \textit{Fig.~\ref{fig:testprocess}}.

Individual system components, such as the blockchain service and smart contracts, are developed in individual repositories maintained with versioning tool of source code. Automated testing strategy is also adopted with different types of functional system tests, namely unit tests against individual functional definitions, static analysis on smart contracts, and flow tests involving different functions per specific operational logic, are also included. Integration tests are also covered between different system components, such as blockchain service and app-backend service on its integration involving endpoint invocations of the blockchain service instance. Good test coverage of blockchain service could demonstrate the assurance on functions developed in different software components do define and execute as expected according to those operations explained in the system implementation of dNAS described in \cite{dnas}.

\subsection{Performance Testing of dNAS}
Performance testing is also included for different system components of dNAS, such as the smart contracts, blockchain network and blockchain service, so as to examine whether the proposed dNAS could fulfil the non-functional requirements set out in \cite{towardblockchain}. There are three key performance metrics on dNAS, including \emph{averaged gas spent per individual method} of the deployed smart contracts, the \emph{averaged transaction per second} (TPS) the chosen enterprise blockchain protocols could process, and the total computation time needed for different operations in data processing phase of system implementation.

\begin{figure}[h]
    \centering
    \captionsetup{justification=centering}
    \includegraphics[width=0.5\textwidth]{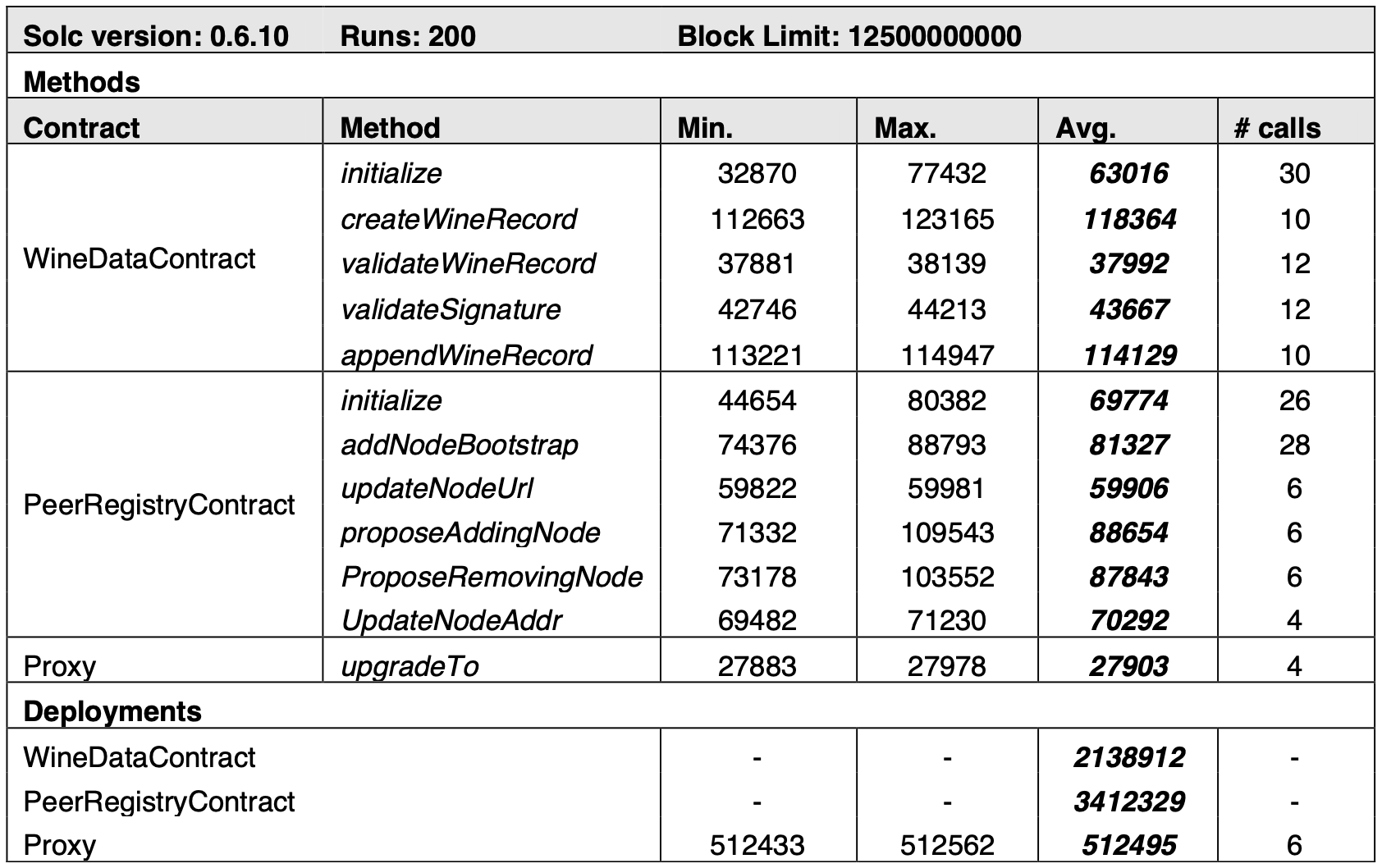}
    \caption{\textit{Average Gas Spent Per Smart Contract Method and Deployment}}
    \label{fig:averagegasspent}
\end{figure}

The average gas spent per method of the deployed smart contracts defined in dNAS, is determined by a load testing of having \emph{200} runs on every method. Deviation on every run per smart contract method is expected owing to the variable on the different states every method needed to process for every run. It could further explain why \emph{averaged gas spent} is a vital metric one should be taken into consideration, so as to better determine the values of initial block gas limit defined in the genesis file of the chosen enterprise blockchain protocols adopted in dNAS, the minimum target of block gas limit on future mined blocks as well as the options on hardware components running the blockchain nodes, such as the Amazon EC2 instance types. The averaged gas spent of every smart contract method alongside the averaged gas spent of individual smart contract deployment process for smart contracts, based on the version of the Solidity compiler at \emph{0.6.10}, are listed in \textit{Fig.~\ref{fig:averagegasspent}}.

On-chain methods of the deployed smart contracts are those with state transitioned, such as the "\emph{createWineRecord}" with the most gas spent amongst all on-chain methods, averaged at \emph{118,364}. Given the targeted minimum block gas limit was set at \emph{12,500,000,000}, it enables the possibility of having multiple transactions related to different methods, validated and packed in the same block as long as the total gas limit is not exceeding predefined block gas limit.

\begin{figure}[h]
    \centering
    \captionsetup{justification=centering}
    \includegraphics[width=0.5\textwidth]{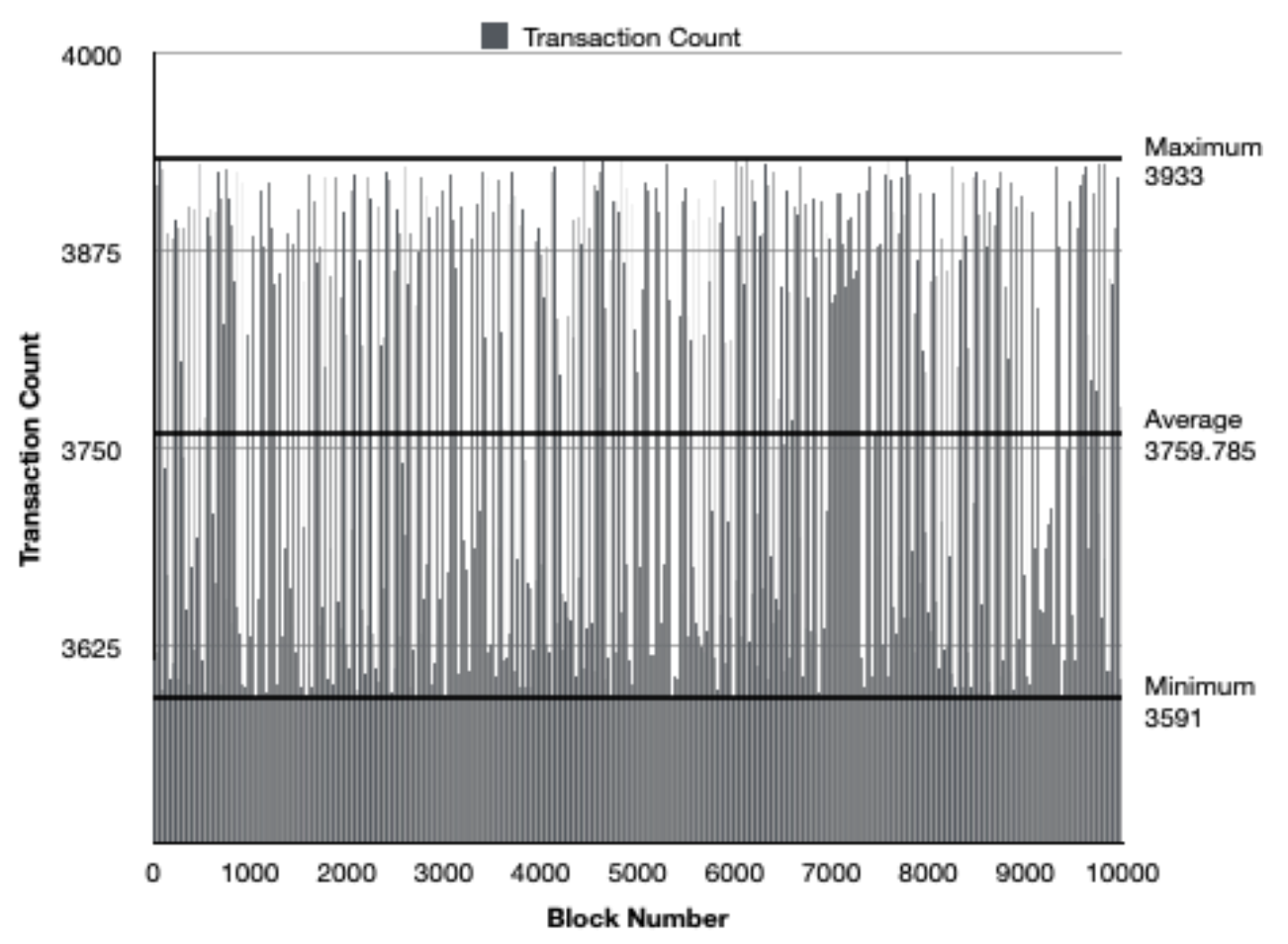}
    \caption{\textit{Distribution of Transaction Counts Over 10,000 Blocks on dNAS Blockchain Network}}
    \label{fig:transactioncount}
\end{figure}

Given the fact that "\emph{createWineRecord}" spent most averaged gas, which is averaged at \emph{118,364}, amongst all the smart contract methods of dNAS as listed in \textit{Fig.~\ref{fig:averagegasspent}}, it is the preferred smart contract method to determine an averaged transaction per second (TPS), based on \emph{10,000} 5-second blocks mined in a 4-node PoA Clique blockchain network under which each node is deployed with a "\emph{t2.2xlarge}" of AWS EC2 instance. \textit{Fig.~\ref{fig:transactioncount}} shows the distribution of transactions validated per block across a 10000-block duration.

According the results shown in \textit{Fig.~\ref{fig:transactioncount}}, the peak and averaged number of transactions validated per \emph{5-second} block are \emph{3,933} and \emph{3,759}, and therefore the peak and averaged transaction per second would then be \emph{786.6 TPS} and \emph{751.8 TPS} respectively. There are in total of \emph{38,565,382} transactions validated across the first \emph{10,000} blocks mined in the network. The results gathered, based on the preferred setting of the blockchain network developed for dNAS, show that non-functional requirements about system scalability set out in \cite{towardblockchain}, of which these requirements requires dNAS should be able to process \emph{one-million} wine record without adversely affecting other operational parameters, and the blockchain network developed as part of dNAS should be able to process \emph{at least 500 transactions per second}, are fulfilled.

\begin{figure}[h]
    \centering
    \captionsetup{justification=centering}
    \includegraphics[width=0.5\textwidth]{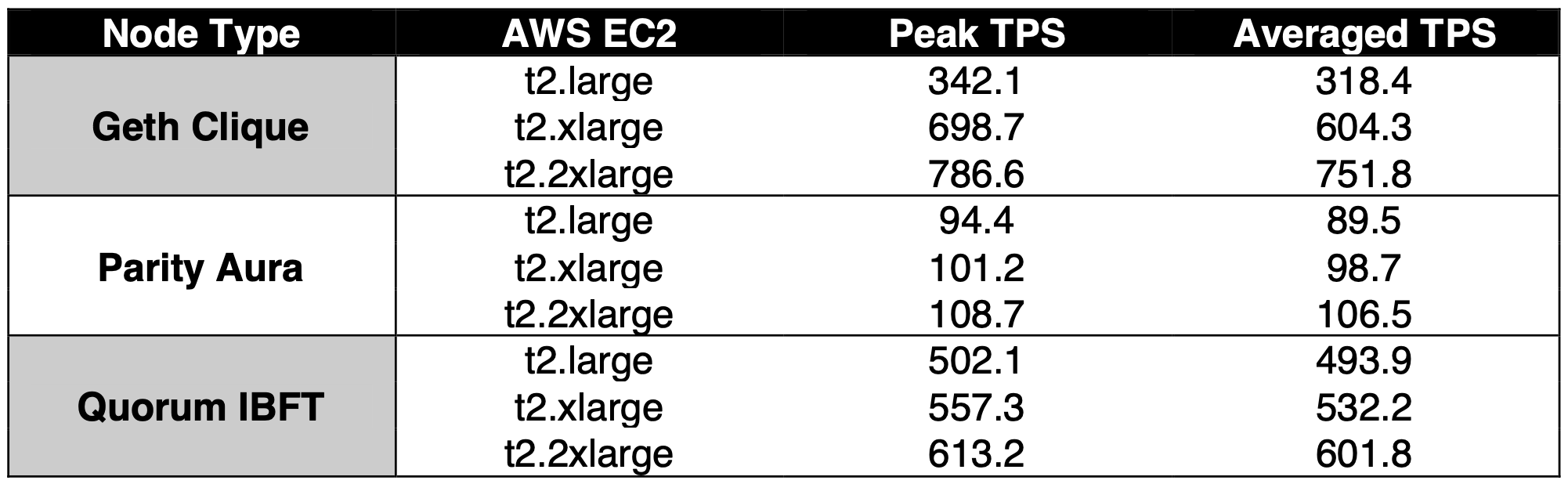}
    \caption{\textit{Transaction Per Second on Different Node Types Deployed on Different types of EC2 Instances}}
    \label{fig:tpsdifferentnode}
\end{figure}

The experiment of determining the transaction per second over \emph{10,000 blocks}, for a \emph{4-node} network has further extended to examine how TPS could change with different types of EVM nodes of Ethereum-based enterprise blockchain protocols, deployed with a variety of choices on AWS EC2 instances with the metric listed in \textit{Fig.~\ref{fig:tpsdifferentnode}}.

According to the comparison and reasoning on the preferred enterprise blockchain protocol for dNAS included in the analysis of consensus protocols covered in the chapter of system analysis on dNAS in this research, \emph{Go-Ethereum Clique} network is of the best performance in terms of scalability, across different types of AWS EC2 instances, compared to different enterprise blockchain protocols consisted of EVM nodes, such as those running on both Parity and Quorum.

\begin{figure}[h]
    \centering
    \captionsetup{justification=centering}
    \includegraphics[width=0.5\textwidth]{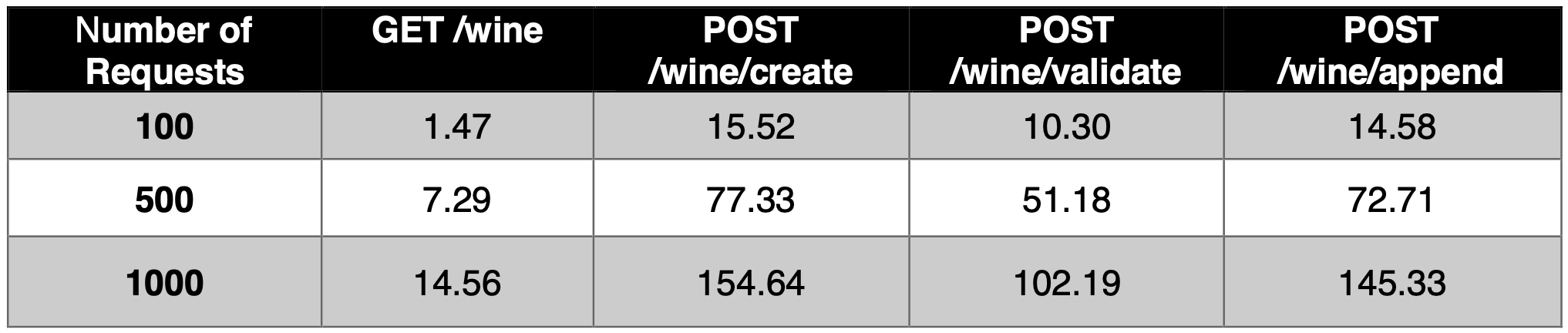}
    \caption{\textit{Computation time (in seconds) of Different Endpoints in Blockchain Service}}
    \label{fig:computationtimednas}
\end{figure}

Load testing activities are also performed to determine the computation time required, with different batches of request sending to different endpoints, related to data processing operations, provided in the blockchain service. The computation time for different sets of load testing, applied on a blockchain service instance of dNAS, is listed in \textit{Fig.~\ref{fig:computationtimednas}}.

\begin{figure}[h]
    \centering
    \captionsetup{justification=centering}
    \includegraphics[width=0.5\textwidth]{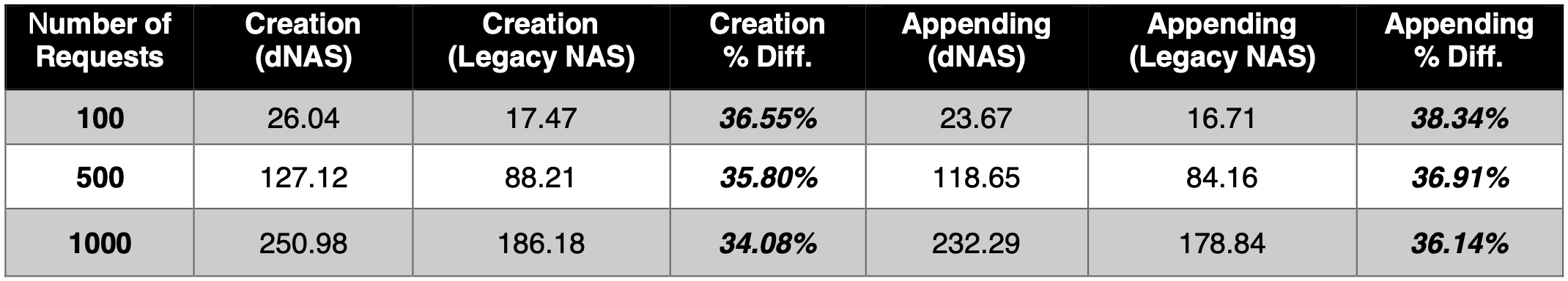}
    \caption{\textit{Computation time (in seconds) of End-To-End Data Processing Operations 
Between dNAS and Legacy NAS}}
    \label{fig:computationtimelegacy}
\end{figure}

The end-to-end load testing is also covered in different end-to-end data processing operations performed, by an instance of dNAS, such as the wine record creation and wine record appending operation, benchmarked with its counterparts performed by the legacy NAS to examine what decentralization could mean to different data processing operations of a supply chain application, in terms of the scalability as listed in \textit{Fig.~\ref{fig:computationtimelegacy}}. The wine record validation operation of dNAS is not included since there is no such operation implemented in the legacy NAS.

According to the results of \emph{1,000} standard requests of both creation and appending operations sent to both dNAS and the legacy NAS, as listed in \textit{Fig.~\ref{fig:computationtimelegacy}}, dNAS would roughly take \emph{34.08\%} and \emph{36.14\%} longer, in terms of the computation time, of creation and appending operation respectively, compared with these performed in the legacy NAS, given the decentralized processes and system components introduced in dNAS.

\section{Discussion of Research Results}
With the system analyses and system testing of dNAS already demonstrated and elaborated, it is important to evaluate, in different perspectives, if these aspects and dNAS itself are consistent with what were set out in the research objectives, on whether dNAS and other decentralized solutions developed basing on enterprise blockchain protocols, and what the system requirements were set out in \cite{towardblockchain}, with the discussion of results according to the findings gathered from a series of system testing activities and system analyses performed against dNAS.

\subsection{The Scalable Decentralized Wine Record Management}
It is expected that decentralized systems are generally less scalable than its centralized counterparts owing to the fact that consensus is needed for every state changed on the data. dNAS is also appeared to be less scalable than the legacy NAS in different operations, such as wine record creation and appending for which the computation time of both with dNAS are \emph{34.08\%} (250.98 seconds per 1,000 requests of wine record creation to be processed) and \emph{36.14\%} (232.29 seconds per 1,000 requests of wine record appending to be processed) more than those operations performed with the centralized legacy NAS. Given the extra decentralized processes, which involve blockchain service, IPFS network, key vault service of dNAS, from the point of invoking endpoints of blockchain service all the way to the blockchain network, the \emph{one-third} extra computation time required only for these decentralized processes, does look reasonably swift if comparing with benefits brought by decentralization, such as the strengthened data integrity and improved system security with distributed instances enabling individual nodes along the supply chain collaborating to combat product counterfeits.

The fact is that computation time of decentralized on-chain processes is always a surplus to the computation time taken for off-chain processes performed in centralized system components inherited from the legacy NAS, such as the app-backend service. In other words, the \emph{one-third} extra computation time, of dNAS compared with the legacy NAS, is contributed only by the decentralized processes, which is of better performance, given the fact that the same wine records are processed on-chain in a distributed and decentralized way. The blockchain service of dNAS is able to handle an average of \emph{6.47 requests per second} (154.64 seconds per 1,000 requests) for wine record creation operation, and an average of \emph{6.88 requests per second} (145.33 seconds per 1,000 requests) for wine appending operation, while the blockchain service can also handle \emph{9.79 requests of wine record validation process per second} (102.19 seconds per 1,000 requests) involving the \emph{three-layered} validation steps (validating an IPFS hash of wine records on-chain, validating wine record subsets stored on IPFS network, and validating wine record subsets retrieved against its counterpart stored in the off-chain database of the app-backend service).

There are two reasons contributed to such a good performance on decentralized system components, which are (1) the selection of the enterprise blockchain protocol, for this case the EVM-based consensus protocol such as the PoA Clique, targeted on good scalability to get started for the blockchain network of dNAS, and (2) the optimized on-chain data storage. The former refers to the selection of permissioned Proof-of-Authority implementation on Go-Ethereum as the preferred consensus algorithm for the enterprise blockchain implementation, instead of a public permission-less Ethereum network with more focus on the decentralization over scalability, for the blockchain network under which the network throughput outdid the \emph{500 TPS} set in non-functional requirements, further to \emph{786.6 TPS} with blockchain node hosted with \emph{t2.2xlarge} of AWS EC2 instance. The enterprise Proof-of-Authority protocol is also proved to be the most scalable option amongst the available enterprise blockchain consensus protocols explained in this research, based on the setting of dNAS, amongst networks with EVM-based blockchain nodes such as the Quorum and Parity Aura. The latter refers to the design decision made that only IPFS content hash representing specific wine record subset is stored on-chain for specific wine identifier, instead of the full version of wine record which would in all likelihood spend \emph{3,060,000} of gas amount, based on the calculation performed in \textit{Appendix~\ref{a1}}, for every state-transitioned transaction on the wine record. The averaged gas amount spent for transactions of wine record creation and appending in the setting of dNAS, are merely \emph{118,364} and \emph{114,129} respectively, according to \textit{Fig.~\ref{fig:averagegasspent}}, which could further be lowered with more optimization mechanisms applied to methods of the smart contracts. The relatively low gas spent on methods of the smart contracts implies that less computation time and gas spent is required per transaction and so more transactions could be packed into a mined blocked as long as the sum of gas spent of transactions to be packed is less than the predefined block gas limit.

Another layer of Merkle proof mechanism, in addition to the one already adopted on the blockchain node regarding transaction states, on the off-chain storage of wine records was once considered. Nonetheless, given the IPFS mechanism is not only to obtain a root hash for a wine record, which is indeed adding another layer of security to the system and to data integrity of the wine record, the IPFS approach is therefore preferred in dNAS. The scalability of dNAS as a whole could be further enhanced with multi-threaded processes introduced to individual system components, such as the app-backend service and blockchain service. More optimized gas spent for individual methods of the deployed smart contracts with more transactions included in a mined block, and other options of non-EVM permissioned consensus algorithms of enterprise blockchain protocols targeted on high scalability, such as the \emph{leader-based} Hyperledger Fabric, could also enhance the scalability of dNAS.

\subsection{Improved Data Integrity}
The \emph{three-layered} data storage and validation on wine records across the off-chain database of the app-backend service, IPFS network and the blockchain network, have proved that data integrity on wine records processed is enhanced under the setting of dNAS. Except the wine record creation operation, any wine record to be processed is required undergoing off-chain validation processes against the full version of wine records, stored in the database dedicated to the app-backend service, and on-chain validation processes against wine record subsets stored on the IPFS network and other data stored on-chain, such as the content hash of IPFS, the write count and the public address related to specific wine identifier, before any operation, such as the wine record appending operation with state transitions on wine records could be executed. New entry of supply chain data and transaction data, with data fields like transaction hash and block number, are then created and added to wine records in database with the content hash of IPFS also updated as the version of wine record subset has been updated.

The on-chain and off-chain validation processes described in the wine record validation operation, presented in the system implementation of dNAS, were in place to ensure data integrity and to prevent any attempted attacks, such as the cloning attack on NFC tags, modification attacks in case the wine identifier and the signature stored in NFC tags are inconsistent with that stored in the database or on-chain storage, and reapplication attacks in case the read count and write count are inconsistent to its counterparts stored off-chain and on-chain respectively. With a certain degree of decentralization attained with both IPFS network and blockchain network, data integrity of processed wine records is further improved, owing to the fact that any state change on a specific wine record with its transaction will now need to be validated with consensus reached on the network. The immutability of transaction states related to state transitions of specific wine record operations would mean that any state change processed on the network could be referred and queried in any connected block explorer, based on individual transaction hashes and block numbers specified.

\subsection{Strengthened Security Considerations}
Given the security considerations deployed to different validation processes of the wine record validation operation to improve data integrity, and according to the result from the threat analysis performed on the legacy NAS, as detailed in \cite{towardblockchain}, which in turn transformed into system requirements for the development of dNAS, a variety of security attacks existed in the legacy NAS have no longer been found in dNAS or error is thrown if those attacks are detected before a state transition could be completed on a specific wine record. The security model on data integrity of dNAS is also proved to be able to function beyond the retailer points, as long as post-purchase wine consumers of consumer-to-consumer market are also registered as registered nodes of dNAS.

There are also various security considerations deployed to different system components of dNAS; for instance, password protection was applied to any tag-reading and tag-writing process on NFC tags given this newly supported feature on the selected model of NFC tags - \emph{NTAG 216 Ferrite}. Some validation processes, as described in the wine record validation operation described in \cite{dnas}, also involve signatures and signing processes. The distributed key vault service instance, assigned to every consortium member alongside its blockchain service instance, is also included to dNAS as a key management module to store and manage key secrets, such as secrets of Ethereum wallet for specific consortium member. Regarding security considerations applied to the deployed smart contracts, as described in the working prototype of dNAS detailed in \cite{dnas}, multiple "\emph{require}" statements are developed and included in different methods to prevent potential attacks, such as reapplication attacks in which the on-chain write count and its counterpart off-chain do not match. Design patterns with role-restriction concept of the smart contract, in the form of "\emph{modifier}", is also introduced, so as to enable access authorization to different methods of the deployed smart contracts.

\subsection{High Availability of System Functionalities}
Ensuring high-level operational performance of different system components to maintain system functionalities on different wine record operations is key to the system implementation of dNAS. With the focus of system security and data integrity on wine record now applied to decentralized system components of dNAS, \emph{availability} of states stored and decentralized system components, such as the blockchain service, blockchain nodes and the IPFS nodes, becomes more significant to the overall availability of system functionalities. 

Given the decentralized nature of dNAS, availability and resilience on the data and states, stored on the blockchain network and IPFS network, are assured and even be enhanced with increasing number of consortium members with their assigned nodes running on both blockchain network and IPFS network, owing to the fact that each node of these networks keeps the copy of the states stored in the persistent volume dedicated to the distributed nodes. The availability of the blockchain network would also be enhanced with more consortium members with their blockchain nodes running on the blockchain network in which the availability could be preserved as long as there is at least a blockchain node running on the network.

The persistent volume storage, assigned to each node instance running on the blockchain network of dNAS to store the blockchain states and the individual chain data, is also proved to be contributed to a faster synchronization and data recovery on states to new blockchain nodes connected to the network as part of the on-boarding process of the consortium. It is due to the fact that failed blockchain nodes can now synchronize from where it left instead of synchronizing from the start – \emph{the first block}. This will then assure the availability of blockchain nodes and the blockchain network as a whole, as failed blockchain nodes could be reconnected to synchronize and process transactions sent to the network immediately. 

The smart contract upgradeability with the proxy pattern is demonstrated to enhance the availability of the blockchain service, for which it does not need to be brought down the individual blockchain service instances every time a smart contract is upgraded with a new contract address. The availability of data stored on-chain will also be preserved with such pattern adopted to upgrade smart contracts with changes on the constituent functional methods. Though the off-chain database and the app-backend service itself are not made distributed under the chosen hybrid approach, the availability will further be enhanced with the fully decentralized approach under which every consortium member will be hosting their own instance of literally every system component in dNAS.

\subsection{Manageable System Integration Model}
Regarding the integration between system components of legacy NAS, such as app-backend service, and decentralized system components including the blockchain network, the system architecture proposed for dNAS has made it rather straightforward for which the data model of wine records originally defined in the legacy NAS is updated with new data fields, such as "\emph{transcation_hash}" and "\emph{blockchain_number}" under "\emph{transaction_data}" according to the sample data model of wine record applied to dNAS as listed in \textit{Appendix~\ref{a1}}. With the new data model of wine record, the app-backend service inherited from the legacy NAS can integrate with decentralized system components of dNAS by including decentralized processes in original wine record operations via predefined functions of invoking the endpoints of blockchain service.

The hybrid approach of system integration model was applied to dNAS under which instances dNAS system components, including the blockchain nodes, IPFS nodes and distributed key vault service, are hosted by consortium and assigned to individual consortium member with secret keys owned by the consortium member and stored in the key vault. The hybrid approach has demonstrated to provide another layer of indirection, allowing consortium members to safely manage their own keys, the backup of the wine records in case anything unexpected goes wrong with the blockchain network, IPFS network or any system components of dNAS, the consortium could still be able to act as a fail-over and manage requests from consortium members. It is understood that not every consortium member would possess the in-house technical capacity to maintain its own instance of dNAS. Such approach of system integration model is proved to be appropriate to help promote the adoption from different nodes along the supply chain of wine industry and for different stakeholders of the industry to collaborate for good to help improve the worsening situation of product counterfeits, with process integrity also conserved.

\section{Limitations and Concerns}
With dNAS is developed and evaluated with a variety of system tests and analysis, limitations and concerns across different system components of dNAS are also identified.

\subsection{Degree of Decentralization}
Though dNAS is proposed and developed to decentralize the legacy NAS using blockchain technology, dNAS is indeed built around enterprise blockchain protocols and a concept of enterprise consortium consisted of registered nodes along the supply chain. Every consortium member is assigned with a blockchain node running on the blockchain network implemented with enterprise blockchain protocols such as PoA or IBFT as the chosen consensus algorithm. Due to the fact that dNAS is developed with the proposed hybrid approach, every instance of dNAS with any system role alongside their distributed nodes of the blockchain network and the IPFS network, are hosted with the enterprise consortium maintained and operated by the consortium administrator.

The decentralized model of dNAS is substantiated with its blockchain network and IPFS network. Implementations with both networks do provide certain degree of decentralization when it comes to transactions being validated and packed in a block on-chain with the corresponding methods of the deployed smart contracts invoked as well as proposing state changes on the consortium registry and the blockchain network. However, decentralization around smart contract management and deployment, management of dNAS instances and management of the distributed network protocols, is limited as these are solely handled by the consortium administrator of dNAS.

\subsection{Scalability Concerns}
\begin{figure}[h]
    \centering
    \captionsetup{justification=centering}
    \includegraphics[width=0.15\textwidth]{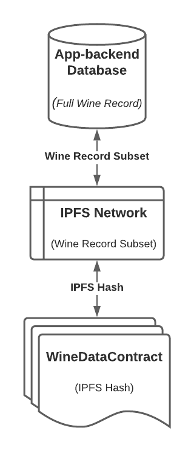}
    \caption{\textit{Data Storage of dNAS}}
    \label{fig:datastorage}
\end{figure}

Scalability can somehow contradict with decentralization, according to the results of performance testing, for which dNAS generally takes longer computation time for a same set of operations performed by the legacy NAS. It is understandable dNAS is less scalable than the legacy NAS, given the advantages of decentralization applied to dNAS with consensus reached for every state change made to a wine record. There are concerns and limitations on scalability identified in dNAS.

dNAS provides three layers to store the representations of a wine record or a full version of the wine record itself as depicted in \textit{Fig.~\ref{fig:datastorage}}. When the number and size of wine records growing with more wine products circulating in the supply chain, longer computation time is definitely expected for wine record operations with more computation resources spent on the data processing on wine records stored in the database app-backend service.

Despite the averaged gas spent of every method and the deployment of the smart contracts are far lower than the block gas limit predefined, the averaged gas spent of smart contracts could still be optimized with lower gas spent. It implies that there are less logical steps needed and so more scalable for a transaction to be executed, validated and packed in a block with same computation resources assigned to the blockchain nodes. Repeated and similar design pattern should be minimized and reused in the smart contract development, so as to optimize the averaged gas spent per method. 

The blockchain will grow bigger when time goes on with more transactions validated and packed in a block mined, not to mention a new block will be created every \emph{5 seconds} for the blockchain network, configured and developed in dNAS. It could take fairly long period of time for a new blockchain node, assigned to the new consortium member, to synchronize with other blockchain nodes to get to the latest global states of the blockchain network. The long synchronization time would hinder the user experience for new consortium members using dNAS when the size of blockchain is too bulk. The requests to the app-backend service and the blockchain service itself are currently handled sequentially for which a new request will be processed only if the previous one is completed or properly handled. The single-threaded handling of these services, involved in any end-to-end wine record operations, would hinder the scalability of dNAS as a whole.

\subsection{Potential Security Vulnerabilities}
dNAS is believed to have protection deployed over cloning attacks, modifications and reapplication attacks, to wine records and its NFC tags. There are specific security vulnerabilities also identified across different constituent system components of dNAS. Every registered node of the enterprise consortium developed in dNAS, is normally assigned with blockchain node and an account, such as Ethereum account if EVN-based enterprise blockchain protocols are applied, of which a key pair is handed over to a registered node for storage and management, with their own instance of key vault service. Though a security authentication layer has already been added on top of the key vault service whenever the corresponding blockchain service instance, owned by the same registered node, retrieves the key pair. The "\emph{Signer}", based on the key pair, is instantiated to validate and send transactions with the local blockchain node via a chosen blockchain client protocol. The same secret private key is also retrieved and signed on the identifiers of wine products, devices and NFC tags, whenever there is a state of the wine record requires to be transitioned with a signature produced, stored on the corresponding NFC tag and on-chain. The same key pair is retrieved and used over and over again without a concept of key rotation, and it is possible that the key pair could be compromised and hence \emph{the aforementioned attacks could still be made possible} to create vulnerabilities and threats to dNAS.

Following the compromised key secrets, \emph{distributed denial-of-service} (DDoS) could also be made possible, with the presence of extremely high gas limit. As long as the blockchain node is hosted with enough computation resource, it is possible to spam the blockchain network with huge amount of transactions to be processed. The denial-of-service could also be performed by any malicious registered node though they are part of the consortium, wine industry and the wider supply chain industry. While for wine record subsets, in \emph{JSON format}, stored on IPFS network with the content hash obtained, anyone can interpret and retrieve the wine record subsets if the content hash is supplied. The wine record subset stored on IPFS is not encrypted because the wine record subset is in all likelihood not sensitive and only open to consortium members. With their IPFS nodes running on a private network, encryption on wine record subsets would ensure the security and access on data itself to be shared to related supply chain participants. The publicity of the smart contract source code could also cause security vulnerabilities. Unlike the source code of different system components in the legacy NAS which has the option to have its code base open-sourced or completely privatised, smart contract code of dNAS is always easily accessible by the nodes running on the blockchain network, and so malicious consortium members running blockchain nodes could look for human induced vulnerabilities if any method of the deployed smart contracts is not implemented correctly.

\subsection{Privacy Concern}
As discussed in the potential security vulnerability, the lack of key rotation mechanism in dNAS would also mean that the same public address is possible to be mapped to an actual registered node. The system role identifier could further be mapped to the true identity of the representative organization, by other registered nodes who are also consortium members. Although public addresses stored on-chain are already obfuscated with hash functions applied, events will be emitted when methods of the deployed smart contracts are executed, whenever there are new transactions on wine record operations related to the same public addresses. The events are later received by the event listener of every blockchain service instance. With more events emitted with the same public address, it is more likely a specific public address could be mapped to an actual registered node, and so its transaction volume could still be derived by other consortium members which could potentially be its competitors. 

The proposed data structure of the wine record, as demonstrated in \textit{Appendix~\ref{a1}}, will be shared across the supply chain to nodes that are related to the transfer of specific wine products. It is suggested in dNAS that new entry of supply chain data will be added to a wine record whenever there is a successful state-transitioned transfer operation performed with both device data and GPS location data of the tag-interacting process in presence. Similarly, these data fields, with privacy details, will also be logged in the wine record for every unsuccessful validation process. In addition to the public address, these data fields could directly relate to physical entities and cause privacy concern if there is no privacy-preserving technology in place to process these sensitive data fields. If the corresponding NFC tags are not deactivated properly when the respective wine products are consumed, it could possibly lead to a privacy threat based on any non-encrypted or non-obfuscated data fields of wine records stored in NFC tags. Privacy-preserving technologies are required with use cases based on the chosen mechanism to be included in any future upgrade of dNAS regarding the privacy concerns.

\section{Future Opportunities of dNAS}
According to the limitations and concerns identified from the system analyses and a series of system test activities against, a variety of future opportunities on strengthening capability in supply chain anti-counterfeiting and traceability are also suggested and elaborated in this chapter.

\subsection{Production Readiness of dNAS}
With distributed instances of decentralized system components and key management hosted and managed by individual consortium member, dNAS should be developed into a production ready state. The on-boarding process will have to be streamlined so as to integrate with the legacy data pipelines already adopted by the registered nodes in the industry, especially integrating with their enterprise resource planning (ERP) systems to enable automation as well as encouraging user adoption throughout the wine industry and its related supply chain participants.

Continuous integration and continuous delivery pipeline should also be implemented for deployment efforts to different environment clusters for major releases with a production ready state of dNAS to promote adoptions amongst major targeted audience or across a specific supply chain network of wine industry. More analysis on metrics, such as energy consumption, memory utilization and costs on hosting instances, for every wine record operation, should also be performed so as to determine the optimal technical stack for a production version of dNAS.

\subsection{Fully Distributed Approach}
Owing to the fact that dNAS is developed to decentralize the legacy NAS and the proposed hybrid approach of system integration model was introduced and applied to dNAS. The next logical step would be looking to take the incremental effort to fully decentralize the prototype of dNAS, in which every registered node which is also the consortium member, to host their own instance of dNAS as well as the distributed nodes running on the blockchain network and the IPFS network. The fully decentralized approach is about having every consortium member to host their own instance of dNAS. In other words, consortium members are now responsible for key management of their blockchain nodes, any system process involved with signatures, and the interaction between distributed nodes with its designated blockchain service instance. Such approach would require in-house technical capacity of the participating consortium members, to maintain their own dNAS instance.

\begin{figure*}[h]
    \centering
    \captionsetup{justification=centering}
    \includegraphics[width=1\textwidth]{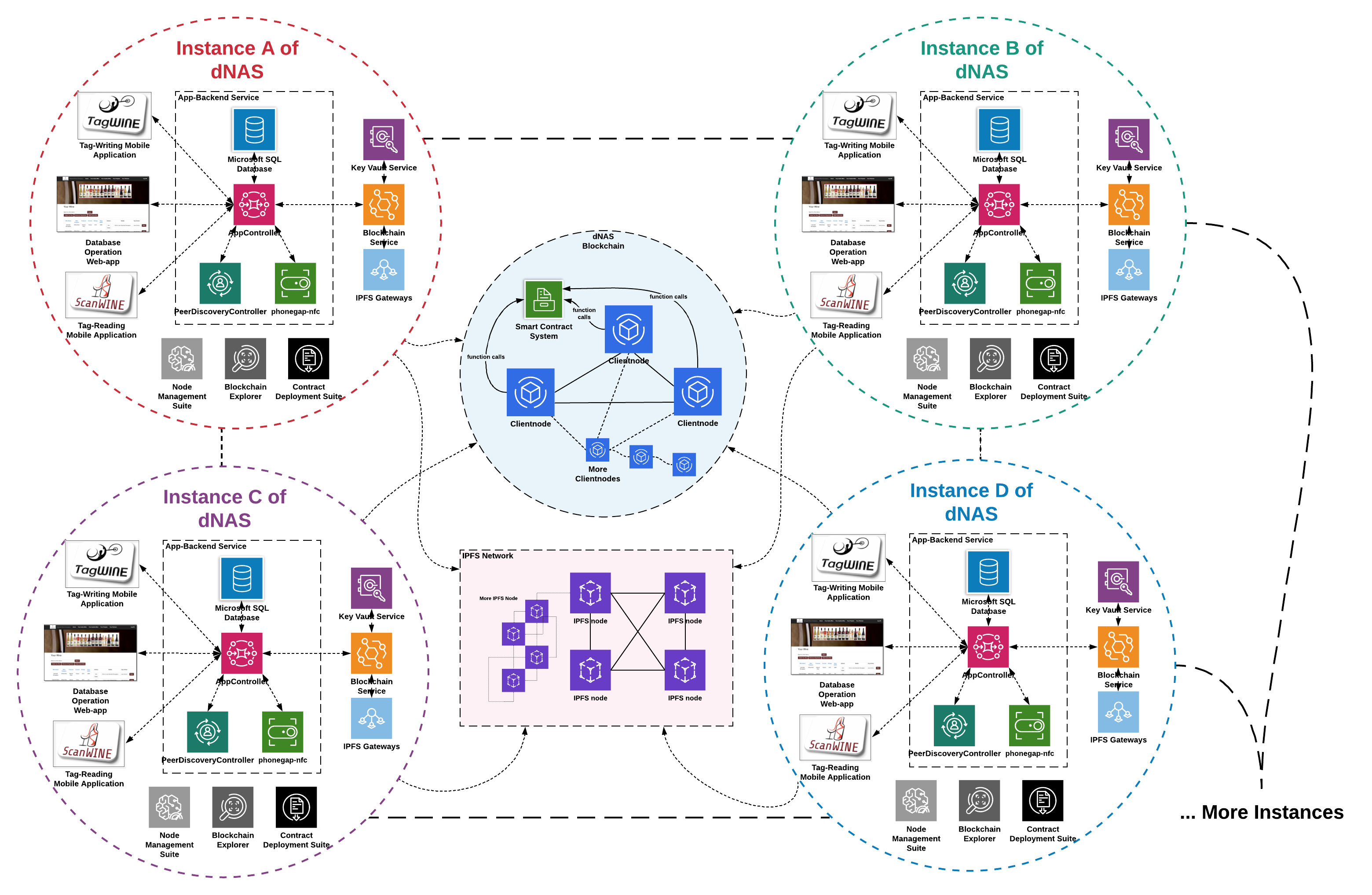}
    \caption{\textit{The Proposed Fully Decentralized Architecture for dNAS}}
    \label{fig:fullydnas}
\end{figure*}

With the advent of distributed dNAS architecture in which every consortium member hosts their own instance of every dNAS system components (registered nodes of wine consumer role would still have to login so as to use the applications and services hosted by the consortium as they are still not part of the consortium), a peer-to-peer controller, such as \emph{PeerDiscoveryController}, will need to be implemented for each instance. Different instances of the \emph{AppController} could therefore synchronize with each other on the states of specific wine product whenever there are operations, such as transfer of wine products or sharing of wine record data, performed along the supply chain. 

\subsection{Adoption of a More Decentralized Enterprise Blockchain Protocol}
The fully distributed approach could also imply that operations of the enterprise consortium could be more decentralized by having more consortium administrators via on-chain governance amongst all the consortium members. The consortium will need to maintain the versioning of individual dNAS components so that every distributed dNAS instance is running with the latest versions of system components. Having such trustful role of consortium administrator completely removed could be an option applied in this approach. Without the consortium administrator, every member in the consortium could perform any operation with their own deployed smart contracts with functionalities shared with other consortium members in different trade groups. It would also imply that every change on the versioning of system components in dNAS will need to go through another set of on-chain governance processes, such as voting process, across the enterprise consortium to reach consensus on different fronts such as the version of smart contracts, version of wine record data model, etc. The concept of a fully distributed approach which should be included in future work of dNAS, is demonstrated in \textit{Fig.~\ref{fig:fullydnas}}.

Before progressing dNAS to the production-ready state, the limitations and concerns detailed should be addressed. On-chain governance mechanism should be introduced on electing the consortium administrators and the number of consortium administrators should also be increased to enhance the degree of decentralization within the consortium and the system implementation as a whole. Other consensus algorithms with a more decentralized way of electing validation leaders validating the transactions per validation group on the blockchain network, could be considered for possible adoption, such as the "\emph{IBFT}" applied to Consensys Quorum. 

Though the \emph{leader-based} consensus algorithm for a blockchain network running with EVM nodes would be roughly 22\% less scalable compared to Proof-of-Authority protocols, for the case of dNAS as evaluated in the system analyses covered in this research. Other mechanisms of improving the scalability could also be introduced, such as having reusable patterns applied to methods of the deployed smart contract and applying the batched pattern operation in the smart contract for which more state-transitioned operations could be included in a transaction and its corresponding block.

\subsection{Upgraded Security Features of dNAS}
Key rotation functionalities should be introduced to dNAS to improve the security aspect on key secrets and dNAS as a whole. The key rotation and key signing could further be implemented with privacy-preserving technique, such as the stealth address \cite{60}, ring signature scheme \cite{61}, multi-signature \cite{62}, or zero-knowledge proof \cite{63}, for more advanced wine record validation concept without a need of signature stored in NFC tags, to improve the privacy model of dNAS. Another layer of network security could be taken into account; for instance, a \emph{virtual private network} (VPN) could be implemented on top of every communication and interaction channel between instances and consortium members in dNAS. A more secured signing process could be performed in \emph{trusted execution environment}, such as Intel SGX for secured computations.

\subsection{Digital Asset Development on dNAS}
The concept of ownership on digital assets with which an on-chain digital twin is created and coupled with product records could be introduced where the notion regarding the proof of ownership could be in place for every transfer operation took place on dNAS. The digital assets could be achieved with the tokenization of on-chain properties related to specific wine records, which could further facilitate the concept of on-chain operations related to digital assets created with functionalities, such as proof of ownership and transfer of ownership, enabled. The concept of digital asset ownership not only strengthens supply chain anti-counterfeiting and traceability of decentralized solutions implemented in supply chain industry, but also opens up more integration opportunities in fields, such as decentralized finance (DeFi) and decentralized e-commerce and exchange platform, for different service sectors in supply chain industry.

\section{Conclusion}
It is concluded that dNAS is a solid example demonstrating how a legacy supply chain anti-counterfeiting and traceability system, for wine industry in this case or for the wider supply chain industry, could be decentralized and reengineered from its centralized architecture which was once hosted and maintained merely by intermediaries or in this case - the winemakers. dNAS is built with enterprise blockchain protocols and around an idea of enterprise consortium consisted of registered nodes along the supply chain of wine industry, balancing degree of decentralization, scalability, security and privacy aspects of such decentralized solution for improved supply chain anti-counterfeiting and traceability, with EVM-based enterprise blockchain protocols, such as PoA and Quorum, chosen appropriately with the preferred consensus algorithms, such as PoA and IBFT, of the blockchain network as well as availability of shared smart contracts with a capability to be upgraded with versions shared amongst consortium members. The hybrid approach of system integration model adopted in dNAS has also proved to be appropriate to help promote the adoption from different roles of wine industry, and for different stakeholders of the industry to collaborate for good to help improve the situation of product counterfeits.

dNAS demonstrated that any state transition on a wine record along the supply chain could be validated and completed automatically, only if, (1) the involved registered nodes and its distributed nodes, such as blockchain nodes and the IPFS nodes, are cryptographically authenticated, (2) the related wine record and the transactions for specific process are cryptographically verified, and (3) the consensus of such state transition is reached with the related transactions packed in a mined block, by blockchain nodes owned by other consortium members. The immutable states of transactions in different blocks are also available on the network and able to be accessed via different tools developed in dNAS, such as the blockchain explorer. The capabilities demonstrated in wine record verification and product provenance with dNAS have confirmed to be strengthened and benefited from a degree of decentralization, under which no more single-point failure and control to be found in dNAS. The resiliency and availability of validated wine records with the history of transitioned states are also enhanced with the persistent volume on states of blockchain and transaction, alongside the wine record data stored in dNAS.

Though dNAS is found to be less scalable than its centralized counterparts as we could expect, the advantage introduced by the concept of decentralization have enabled different nodes along the supply chain of wine industry to work collaboratively with wine records flowing downstream seamlessly with state transitions validated automatically in a decentralized and immutable fashion. dNAS is proved to be not only a verification service but also a notary service, providing a proof of existence for wine records owned by different supply chain participants. All system requirements defined in dNAS were fulfilled with the claimed functionalities confirmed with both functional testing and performance testing. Further qualitative analysis on dNAS as a whole was also performed with limitations and risks identified so as to support rationale of future opportunities on further improving different aspects, such as system security, degree of decentralization and scalability of dNAS. 

As we set out in the research objective, it is confirmed that decentralized solutions, including dNAS, of supply chain anti-counterfeiting and traceability, are more effective and worth than its centralized counterparts of supply chain industry to better improve the situation of counterfeiting attacks in wine industry and even in supply chain industry as a whole. With flexibility on the architecture of decentralized system components and the suggested data structure defined in design decisions made during the system design and development steps, the decentralized solutions and operations proposed in dNAS are indeed industry-agnostic and should be available for any service sectors looking to combat product counterfeit in a wider supply chain industry.

\newpage
\bibliographystyle{ieeetr}

\begin{thebibliography}{10}

\bibitem{1}
OECD and EUIPO, {\em Trade in Counterfeit and Pirated Goods – Mapping The
  Economic Impact}.
\newblock OECD Publishing, 2019.

\bibitem{25}
N.~C.~K. Yiu, ``An nfc-enabled anti-counterfeiting system for wine industry,''
  {\em arXiv preprint arXiv:1601.06372}, 2014.

\bibitem{towardblockchain}
N.~C.~K. Yiu, ``Toward blockchain-enabled supply chain anti-counterfeiting and
  traceability,'' {\em arXiv preprint arXiv:2102.00459}, 2020.

\bibitem{55}
S.~A. Abeyratne and R.~P. Monfared, ``Blockchain ready manufacturing supply
  chain using distributed ledger,'' {\em International Journal of Research in
  Engineering and Technology}, vol.~5, no.~9, pp.~1--10, 2016.

\bibitem{56}
M.~P. Caro, M.~S. Ali, M.~Vecchio, and R.~Giaffreda, ``Blockchain-based
  traceability in agri-food supply chain management: A practical
  implementation,'' in {\em 2018 IoT Vertical and Topical Summit on
  Agriculture-Tuscany (IOT Tuscany)}, pp.~1--4, IEEE, 2018.

\bibitem{57}
K.~Toyoda, P.~T. Mathiopoulos, I.~Sasase, and T.~Ohtsuki, ``A novel
  blockchain-based product ownership management system (poms) for
  anti-counterfeits in the post supply chain,'' {\em IEEE access}, vol.~5,
  pp.~17465--17477, 2017.

\bibitem{58}
Z.~C. Kennedy, D.~E. Stephenson, J.~F. Christ, T.~R. Pope, B.~W. Arey, C.~A.
  Barrett, and M.~G. Warner, ``Enhanced anti-counterfeiting measures for
  additive manufacturing: coupling lanthanide nanomaterial chemical signatures
  with blockchain technology,'' {\em Journal of Materials Chemistry C}, vol.~5,
  no.~37, pp.~9570--9578, 2017.

\bibitem{59}
H.~M. Kim and M.~Laskowski, ``Toward an ontology-driven blockchain design for
  supply-chain provenance,'' {\em Intelligent Systems in Accounting, Finance
  and Management}, vol.~25, no.~1, pp.~18--27, 2018.

\bibitem{dnas}
N.~C.~K. Yiu, ``Decentralizing supply chain anti-counterfeiting systems using
  blockchain technology,'' {\em arXiv preprint arXiv:2102.01456}, 2020.

\bibitem{28}
S.~Nakamoto, ``Bitcoin: A peer-to-peer electronic cash system,'' tech. rep.,
  Manubot, 2008.

\bibitem{30}
E.~Androulaki, A.~Barger, V.~Bortnikov, C.~Cachin, K.~Christidis, A.~De~Caro,
  D.~Enyeart, C.~Ferris, G.~Laventman, Y.~Manevich, {\em et~al.}, ``Hyperledger
  fabric: a distributed operating system for permissioned blockchains,'' in
  {\em Proceedings of the thirteenth EuroSys conference}, pp.~1--15, 2018.

\bibitem{ebc}
A.~Karamchandani, S.~K. Srivastava, and R.~K. Srivastava, ``Perception-based
  model for analyzing the impact of enterprise blockchain adoption on scm in
  the indian service industry,'' {\em International Journal of Information
  Management}, vol.~52, p.~102019, 2020.

\bibitem{supervisedmachinelearning}
I.~M. Cavalcante, E.~M. Frazzon, F.~A. Forcellini, and D.~Ivanov, ``A
  supervised machine learning approach to data-driven simulation of resilient
  supplier selection in digital manufacturing,'' {\em International Journal of
  Information Management}, vol.~49, pp.~86--97, 2019.

\bibitem{42}
J.~Kwon, ``Tendermint: Consensus without mining,'' {\em Draft v. 0.6, fall},
  vol.~1, no.~11, 2014.

\bibitem{43}
E.~Buchman, {\em Tendermint: Byzantine fault tolerance in the age of
  blockchains}.
\newblock PhD thesis, 2016.

\bibitem{60}
N.~T. Courtois and R.~Mercer, ``Stealth address and key management techniques
  in blockchain systems.,'' {\em ICISSP}, vol.~2017, pp.~559--566, 2017.

\bibitem{61}
S.~Noether, ``Ring signature confidential transactions for monero.,'' {\em IACR
  Cryptol. ePrint Arch.}, vol.~2015, p.~1098, 2015.

\bibitem{62}
L.~Harn, ``Group-oriented (t, n) threshold digital signature scheme and digital
  multisignature,'' {\em IEE Proceedings-Computers and Digital Techniques},
  vol.~141, no.~5, pp.~307--313, 1994.

\bibitem{63}
C.~Rackoff and D.~R. Simon, ``Non-interactive zero-knowledge proof of knowledge
  and chosen ciphertext attack,'' in {\em Annual International Cryptology
  Conference}, pp.~433--444, Springer, 1991.

\bibitem{33}
G.~Wood {\em et~al.}, ``Ethereum: A secure decentralised generalised
  transaction ledger,'' {\em Ethereum project yellow paper}, vol.~151,
  no.~2014, pp.~1--32, 2014.

\end{thebibliography}

\begin{IEEEbiography}[{\includegraphics[width=1.1in,height=1.25in,clip,keepaspectratio]{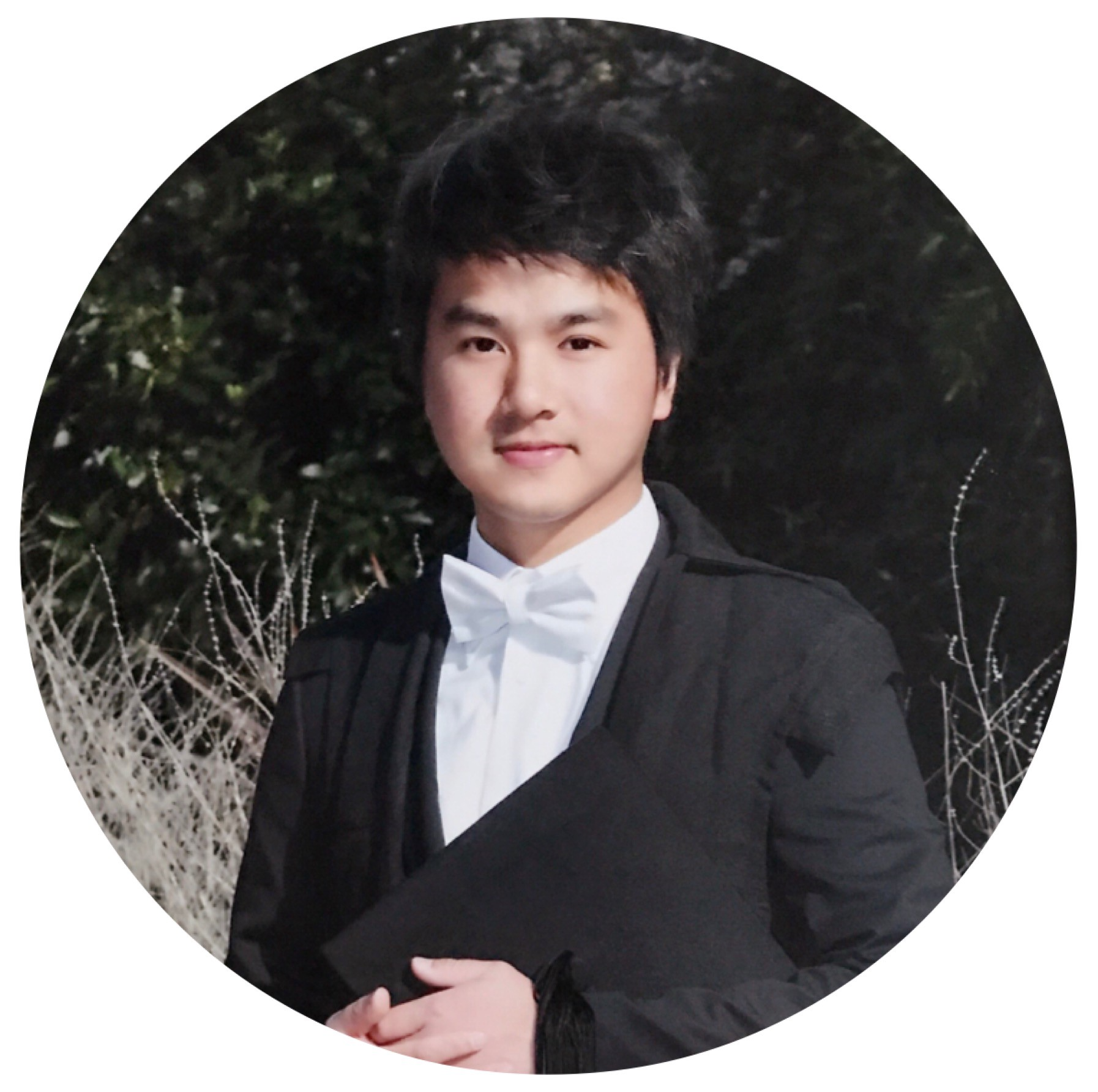}}]{Mr. Neo C.K. Yiu IEEE}
is a computer scientist and software architect specialized in developing decentralized and distributed software solutions for industries. Neo is currently the Lead Software Architect of Blockchain and Cryptography Development at De Beers Group on their end-to-end traceability projects across different value chains with the Tracr™ initiative. Formerly acting as the Director of Technology Development at Oxford Blockchain Society, Neo is currently a board member of the global blockchain advisory board at EC-Council. Neo received his MSc in Computer Science from University of Oxford and BEng in Logistics Engineering and Global Supply Chain Management from The University of Hong Kong.
\end{IEEEbiography}
\vfill

\newpage
\appendices
\section{Data Model Definition of dNAS}
\subsection{Sample Data Model of Wine Record} \label{a1}
\begin{figure}[h]
    \centering
    \captionsetup{justification=centering}
    \includegraphics[width=0.5\textwidth]{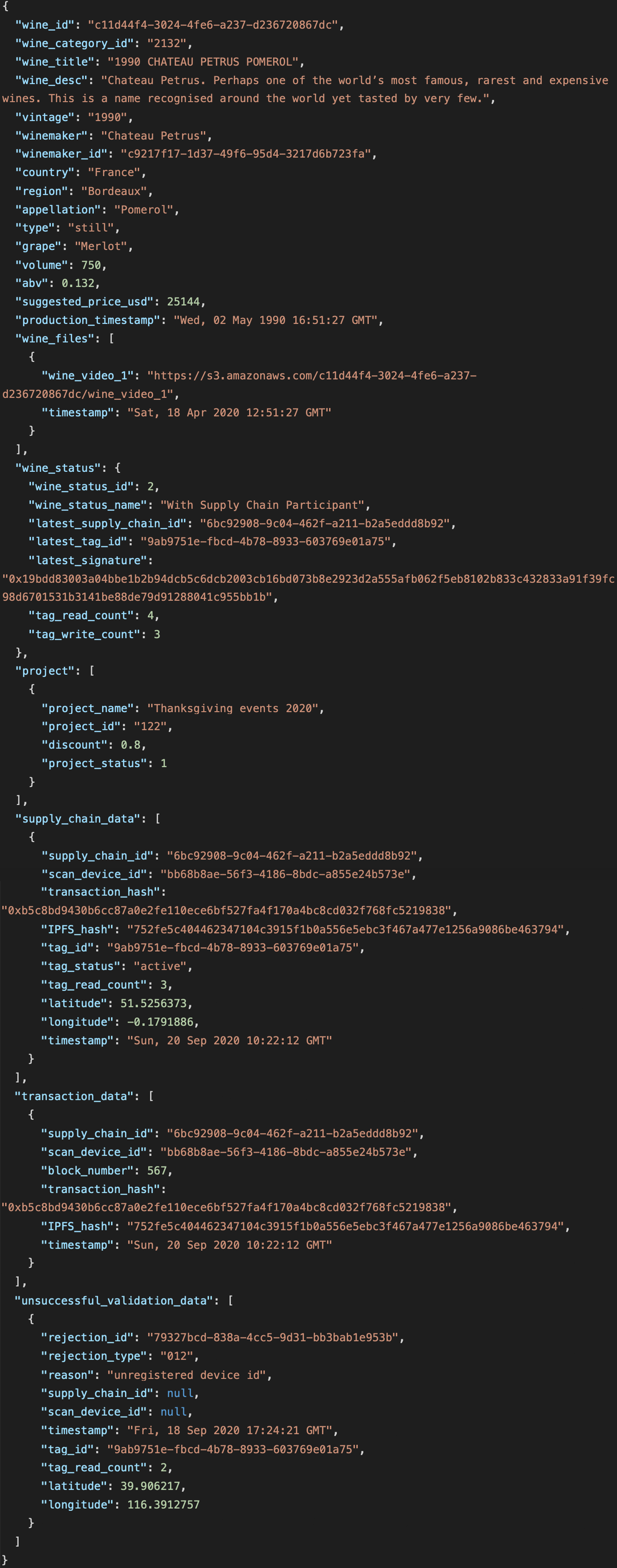}
    \caption{\textit{Sample Data Model of Wine Record}}
    \label{fig:winerecordformat}
\end{figure}

The Sample Data Model of Wine Record defined in dNAS is demonstrated in \textit{Fig.~\ref{fig:winerecordformat}}.

\subsection{Sample Data Components of Wine Record Subset Stored On IPFS} \label{a2}
\begin{figure}[h]
    \centering
    \captionsetup{justification=centering}
    \includegraphics[width=0.5\textwidth]{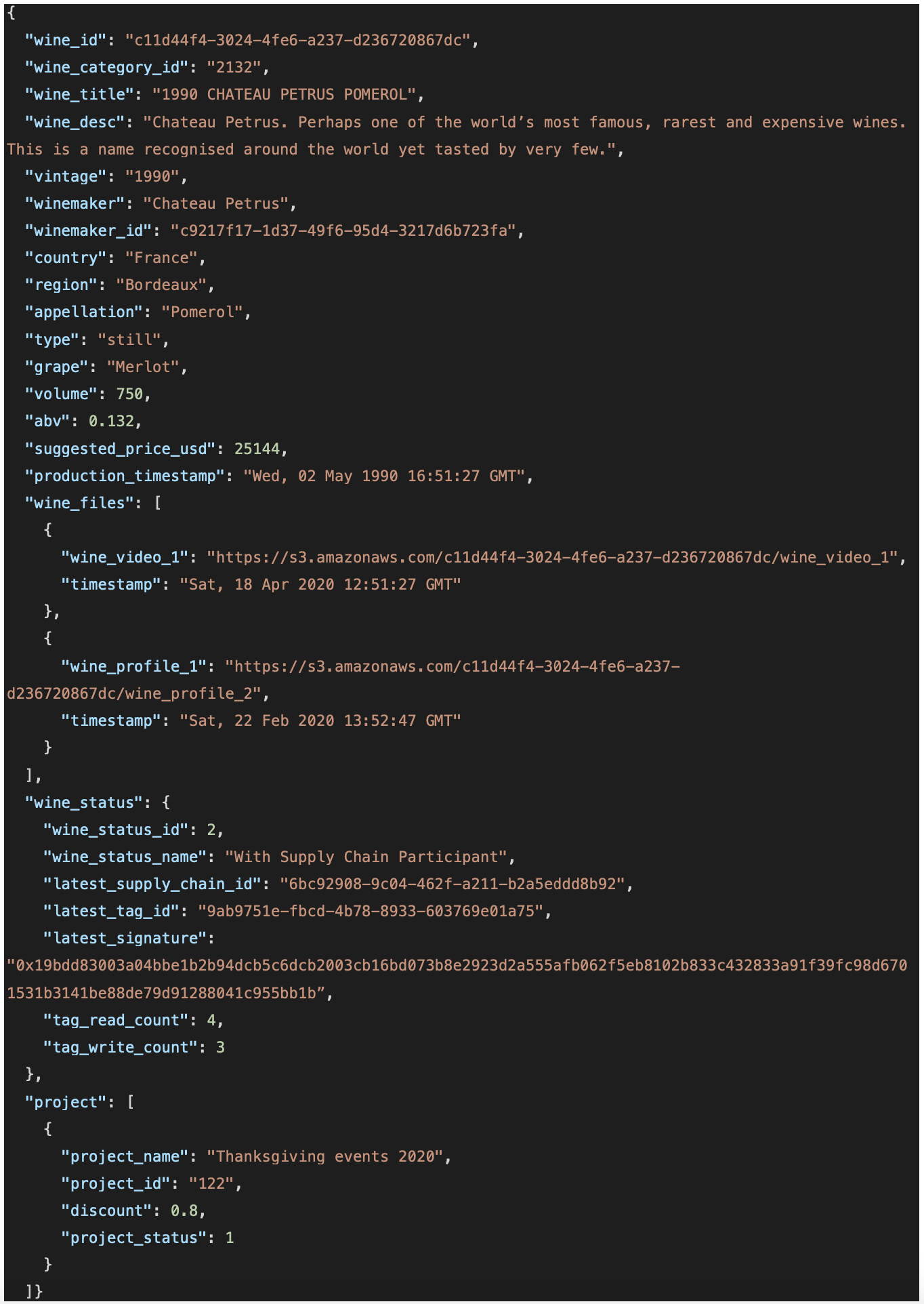}
    \caption{\textit{Sample Data Components of Wine Record Subset Stored On IPFS}}
    \label{fig:ipfsdataformat}
\end{figure}

The Sample Data Components of Wine Record Subset Stored On IPFS defined in dNAS is demonstrated in \textit{Fig.~\ref{fig:ipfsdataformat}}.

\subsection{Data Model Definition of dNAS} \label{a3}
With only minimal changes applied to the data model of wine record, defined in the legacy NAS, data categories such as wine status, supply chain data, transaction data and unsuccessful validation data were updated with more data fields so as to integrate with the decentralized aspects of dNAS alongside strengthened anti-counterfeiting mechanism enabled.

The sample of the proposed data model of wine records applied to dNAS is demonstrated in \textit{Appendix~\ref{a1}}. For instance, a wine product with "\emph{c11d44f4-3024-4fe6-a237-d236720867dc}" as "\emph{wine_id}" has gone through supply chain processes including the first step – production of the wine product which in turn with a wine record created, with individual "\emph{supply_chain_id}" and "\emph{scan_device_id}" of a registered node. In order to have the latest "\emph{wine_status}", both "\emph{supply_chain_data}" and "\emph{transaction_data}" are needed to be updated with data fields such as "\emph{transaction_hash}" and "\emph{block_number}" of validated transactions for specific supply chain operations. As every supply chain operation should involve the operations on NFC tags packaged with individual wine products, data fields such as "\emph{tag_id}", "\emph{tag_read_count}" and "\emph{timestamp}" are logged under "\emph{supply_chain_data}". The "\emph{wine_status}" captures every latest data field, such as "\emph{latest_supply_chain_id}" signalling which supply chain node the wine product is physically located at currently, "\emph{latest_tag_id}" as the NFC tag identifier could be updated only by winemakers if necessary, after returns of related wine products, and "\emph{tag_read_count}" showing number of times a specific NFC tag has been read throughout every operation process performed.

There is also another subset in the proposed schema of wine records, which is critical to dNAS, namely "\emph{unsuccessful_validation_data}" of which any unsuccessful validation operation when scanning the NFC tag at different nodes along the supply chain would be logged, with a "\emph{rejection_id}" in \emph{UUID} and a reason of rejection given. It will also give information such as where and when a specific operation was carried out, which supply chain participant using which of its device to perform such scanning and validating activities, and which NFC tag of a given wine product was attempted to scan. 

The "\emph{wine_pedigree_data}" gives product details about a specific wine product. Only registered nodes of winemaker role could make changes to the "\emph{wine_pedigree_data}" and "\emph{project}" using the database operation web-app as mentioned in the use case analysis. All data fields under "\emph{supply_chain_data}", "\emph{transaction_data}" and "\emph{unsuccessful_validation_data}" could only be updated whenever tag-writing or tag-reading operations are performed on NFC tags along the supply chain via the NFC-enabled \emph{ScanWINE} mobile application and so there is no direct \emph{CRUD} operation on these data fields at any point of an operation performed in any system component of dNAS.

\subsection{Calculation of Gas and Total Cost on the Sample Full Version of Wine Record} \label{a4}
A "\emph{word}" in any EVM-based protocol, such as Ethereum main net, is \emph{32 bytes}, implying that so as to store \emph{2,900 bytes} it would mean storing roughly \emph{91} words' worth of data. Assuming all these words were set to non-zero from zero, and according to the Ethereum yellow paper as explained in \cite{33}:

\begin{itemize}
  \item Gsset 20000 Paid for an SSTORE operation when the storage value is set to non-zero from zero.
  \item Gsreset 5000 Paid for an SSTORE operation when the storage value’s zeroness remains unchanged.
\end{itemize}

This would suggest calling for an \emph{SSTORE} operation \emph{91} times, at \emph{20,000} gas each (the cost to allocate and store a non-zero value). The total gas used to store \emph{2900-byte} size of data would be: 

\[ total\_gas\_used = Gsset * number\_of\_32-byte\_word \]
\[= 20,000 * 91\]
\[= 1,820,000 \]

Each transaction involved a full wine record processed on-chain would mean a whopping amount of \emph{1,820,000} gas used and the cost per transaction, according to the average gas price listed on the Ethereum gas station (https://ethgasstation.info/), which is currently at \emph{60 Gwei}, every transaction involved the whole wine record could cost:

\[ total\_cost = total\_gas\_used * average\_gas\_price \]
\[ = 1820000 * 60 / 10^9 \]
\[ = 0.1092 ETH \]
\[ = \$133.55 (ETH\_price = \$1,223)) \]

It would cost a whopping \emph{\$133.55} to only process a sample wine record on-chain and about \emph{333} seconds and \emph{34} blocks for a transaction to be confirmed, on the main net of Ethereum.

\end{document}